\documentclass[twocolumn]{aastex631}

\usepackage{graphicx}	
\usepackage{amsmath}	
\usepackage{amssymb}	
\usepackage{float}
\usepackage{amsmath}
\usepackage{graphicx}
\usepackage{mathtools}
\usepackage{multirow}
\usepackage{verbatim}
\usepackage{color, colortbl}

\newcommand{\response}[1]{#1}
\newcommand{\responsetwo}[1]{#1} 

\renewcommand{\baselinestretch}{1.1} 

\submitjournal{ApJ}

\shorttitle{Planetesimal Formation by Gravitational Collapse}
\shortauthors{Polak and Klahr}
\graphicspath{{./}{}}

\begin{document}

\title{High Resolution Study of Planetesimal Formation by Gravitational Collapse of Pebble Clouds}

\author[0000-0001-5972-137X]{Brooke Polak}\thanks{Fellow of the International Max Planck Research School for \newline Astronomy and Cosmic Physics at the University of Heidelberg}
\affil{Zentrum f{\"u}r Astronomie der Universit{\"a}t Heidelberg, Institut f{\"u}r Theoretische Astrophysik, Albert-Ueberle-Str. 2, D-69120 Heidelberg, Germany}
\affil{Department of Astrophysics, American Museum of Natural History, 79$^{th}$ Street at Central Park West, New York, NY 10024, USA}

\author[0000-0002-8227-5467]{Hubert Klahr}
\affil{Max-Planck-Institut f{\"u}r Astronomie, K{\"o}nigstuhl 17, D-69117 Heidelberg, Germany}

\begin{abstract}
Planetary embryos are built through the collisional growth of 10-100 km sized objects called planetesimals, a formerly large population of objects, of which asteroids, comets and Kuiper-Belt objects represent the leftovers from planet formation in our solar system.
Here, we follow the paradigm that turbulence created over-dense pebble clouds, which then collapse under their own self-gravity.
We use the multi-physics code GIZMO  to model the pebble cloud density as a continuum, with a polytropic equation of state to account for collisional interactions and capturing the phase transition to a quasi-incompressible "solid" object, i.e.\ a planetesimal in hydrostatic equilibrium.
Thus we study cloud collapse effectively at the resolution of the forming planetesimals, allowing us to derive an initial mass function for planetesimals in relation to the total pebble mass of the collapsing cloud.
The redistribution of angular momentum in the collapsing pebble cloud is the main mechanism leading to multiple fragmentation. The angular momentum of the pebble cloud and thus the centrifugal radius increases with distance to the sun, but the solid size of the forming planetesimals is constant. Therefore we find that with increasing distance to the sun, the number of forming planetesimals per pebble cloud increases. For all distances the formation of binaries occurs within higher hierarchical systems.
The size distribution is top heavy and can be described with a Gaussian distribution of planetesimal mass. For the asteroid belt, we can infer a most likely size of 125 km, all stemming from pebble clouds of equivalent size 152 km.
\end{abstract}



\section{Introduction}

Planetesimals are 1-100 km-sized bodies bound by self-gravity that serve as the building blocks of planets \citep{kokubo10.1093/ptep/pts032}. The planetesimal formation process begins with $\micron-$sized dust particles in circumstellar disks that coagulate into cm-sized pebbles through collisional sticking \citep{safronov}. Then various instabilities in the disk, i.e.\ the streaming instability and vertical shear instability, form pebble traps that collect pebbles into dense, gravitationally bound clouds \citep{Youdin_2005,2007Natur.448.1022J,Cuzzi2008,Cuzzi2010,Hartlep2020,2014prpl.conf..547J,Klahr2018}. Planetesimals are thought to form from these over-dense pebble clouds directly from gravitational collapse \citep{1973ApJ...183.1051G}. These planetesimals continue to grow via mutual collisions, until the largest reach roughly moon size and start to efficiently accrete pebbles \citep{OrmelKlahr2010}. Once this embryo exceeds more than about 10 earth masses, a gravitational collapse of the surrounding gas atmosphere is triggered, followed by run-away gas accretion \citep{Pollack1996}. This eventually leads to forming gas giants. That is the general picture of the core-accretion model of planet formation.

Because planetesimals are the first planet-building objects bound by self-gravity\footnote{One can define the planetesimal mass by asking that the headwind around the planetesimal in the solar nebula can no longer erode the planetesimal. In other words the gravity of the planetesimals must be stronger than the frictional forces acting on the pebbles at the surface of the planetesimal.}, studying their formation and characterizing their initial properties is imperative for understanding the origins of planets. In our solar system, it is believed that asteroids, Kuiper Belt Objects (KBOs), and comets are the planetesimals leftover from the planet formation process. Therefore, observations of these objects give us a glimpse into the properties of the original planetesimal population, albeit after 4.5 billion years of evolution in the solar system.

Many hydrodynamic simulations have confirmed the creation of gravitationally bound pebble clouds from instabilities in disks, and they found the gravitational collapse of these pebble clouds to be an efficient and quick path to planetesimal formation \citep{Johansen_2006,2007Natur.448.1022J,Johansen_2009,Simon_2017,Li_2018}. However, because of the scales involved, hydro simulations of disk instabilities lack the resolution to properly follow the collapse of the pebble clouds into planetesimals. Because of this, dedicated N-body simulations of pebble cloud collapses are needed to accurately model the  planetesimal formation process.

Much of the work focusing on planetesimal formation by gravitational collapse has been done in a series of papers by \citet{Nesvorn__2010,nesvorny2019transneptunian,Nesvorny2021}. They use a collisional N-body code of super-particles representing pebble sub-clouds to model the cloud collapses. In their studies, they were able to reproduce some of the properties of observed KBO binaries. \citet{Nesvorny2021} focuses specifically on pebble clouds formed from the streaming instability (SI) by using clumps formed in SI hydro simulations as their initial conditions. They found that gravitationally collapsing clouds triggered by the SI can indeed produce equal-sized binaries with the observed angular momentum of those found in the Kuiper Belt. They also predict the presence of hierarchical systems with large moons in the Kuiper Belt. They established a relation between the initial distribution of angular momentum in the collapsing cloud and the properties of the binary systems that formed. Their results matching observed KBO properties further confirms the theory that planetesimals are formed by gravitational collapse. Yet, their simulations lack the possible formation of any close binaries and especially contact binaries, due to the resolution limitation in their collision model for super-particle - super-particle interaction and sink-cell approach.

As an alternative prescription for the momentum exchange among super-particles via collisions and avoiding the sink-cell approach, we choose a different approach. We assign a pressure to the individual pebble sub-clouds based on their density, i.e.\ a polytropic equation of state. This allows the formation of hydrostatically stable self-gravitating clumps, at the desired size and density of planetesimals, and simultaneously gives a smooth transition from weak interactions of the dilute pebble clouds to a "solid" or rather quasi-incompressible body. 
Most importantly by using a Riemann solver to treat the interaction of pebble clumps, we have a robust method to study collisions, because for the initially very dilute pebble clouds the assigned pressure is much lower than the ram pressure in a collision. In other words, collisions between dilute pebble cloud clumps are super sonic, and the momentum exchange is proportional to collision velocity and frequency, exactly what one wants to achieve in a statistical collision model.
As a result, we achieve an adaptive resolution model from a large initial cloud to a sub-planetesimal level, without changing the solver en route and without sink particles.

Using this approach, we perform dedicated simulations of gravitationally collapsing pebble clouds to study the planetesimal formation process. Our work builds off the results of \citet{Nesvorny2021} by exploring the effect of disk birthplace on the properties of the formed planetesimals and binaries. Additionally, we investigate the randomness of the initial particle distribution's effect on the outcome of a collapse for one specific location in the disk. The properties we are most interested in are the mass formation efficiency of the collapse, the most massive clump that forms, the total size and mass distribution of planetesimals, and various binary properties such as the size ratio and semi-major axis. 

Instead of using initial conditions from hydro simulations, we start with a uniform density sphere of pebbles motivated by \citet{Klahr_2020,Klahr2021} in rigid and bound rotation around the star. 
\response{They found the diameters of planetesimal to depend primarily on the dust to gas ratio when reaching the local Hill density. The level of streaming instability respectively the diffusion stemming from this instability defines a turbulent Jeans-Mass. For smaller masses, the diffusion is faster than the free fall time and no planetesimal can form. Larger masses than the critical mass can collapse, yet it is argued that the masses are unlikely to exceed the critical mass by a lot, because the buildup of such a massive pebble cloud would take more time than the collapse of the already unstable pebble cloud. Therefor the turbulent “Jeans-Mass” should be the predominant initial condition for pebble cloud collapse. These masses, or respective sizes would correspond to 100 km diameter likewise at 2 AU and 20 AU at the start of planetesimal formation, when the solar nebula contains about $10\%$ of the solar mass, with a tendency towards $10 km$ after a Million years, respectively when the nebula’s mass has decreased to$ 1\%$ of a solar mass. Still all these equivalent sizes characterise the initial mass of the pebble cloud. The present work therefore tries to answer how many objects will form during the collapse process. Some pebbles might escape the collapse entirely and pebbles that do not escape might distribute over a number of objects smaller than the equivalent size.

The actual size distribution therefore needs both works \citep{Klahr_2020,Klahr2021} and the present work. The first sets the total mass of the initial condition and the present work determines the relative mass of the forming fragments.

The today observed size distribution is then set by A: the Jeans mass as function of planetesimal formation time, B: the efficiency of the collapse process, which strongly depends on the heliocentric distance, and finally C: the collisional evolution of the asteroid and Kuiper-Belt populations.
}

\begin{figure*}[ht!]
\centering
    \includegraphics[width=1.0\textwidth]{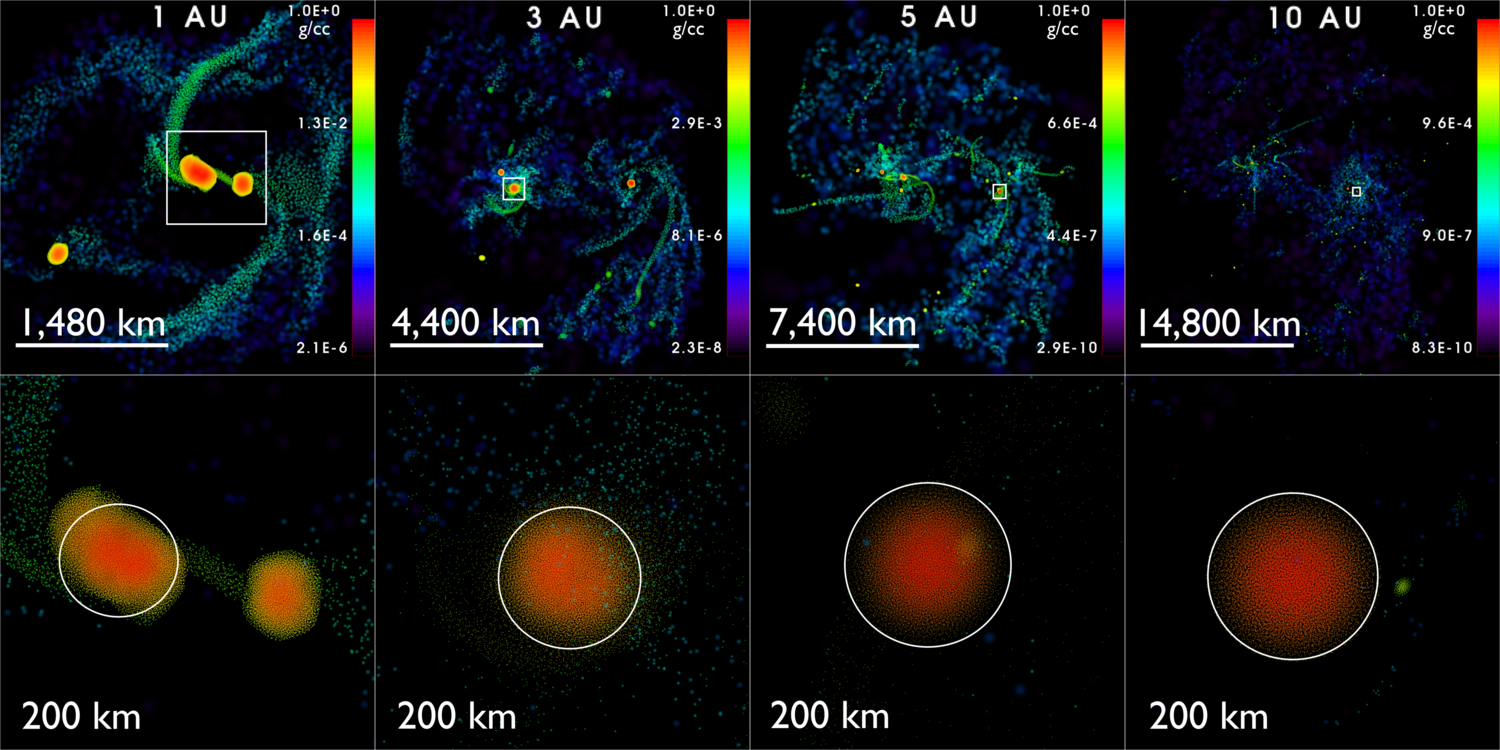}
    \caption{Clumps formed in our simulations at various orbital radii for a 1M$_\odot$ central star. The top panel shows an overview of the entire collapsed cloud with the scale shown in the bottom left. The bottom panel shows a zoomed-in look at the clumps above, indicated by the corresponding white squares above it. The size of formed clumps is \response{<200 km}, indicated by the scales in the bottom panel.}
    \label{fig:AUcollapses}
\end{figure*}

\response{Using this simplified initial condition} gives us complete control over the initial conditions and allows us to isolate trends in planetesimal properties for different cloud locations. We benchmark our model by comparing our results against those for the A12 simulations in \citet{Nesvorny2021} as well as observations of KBOs \citep{2019pdss.data....4J} and asteroids \citep{delbo2017}. Figure \ref{fig:AUcollapses} offers a glimpse of the results from some of our collapse simulations across the solar system. 

This paper is laid out as follows: in Section \ref{section:methods}, we explain the details of our computational model. Section \ref{section:results} goes over our main results, providing a thorough analysis of the planetesimal and binary properties we formed as well as their comparison to observations of asteroids \citep{delbo2017} and KBOs \citep{2019pdss.data....4J} and the results of previous simulations \citep{Nesvorny2021}. We summarize the main takeaways from our results and conclude with our outlook on the future of this work in Section \ref{section:conclusion}.

\section{Methods}
\label{section:methods}

\subsection{Hydro Method}

We performed numerical simulations of the gravitational collapse of pebble clouds using the multi-physics code GIZMO \citep{10.1093/mnras/stv2180}. The predominant feature of the GIZMO code is a novel Lagrangian hydrodynamics method, Meshless Finite Mass (MFM). MFM works by initially partitioning the domain into fluid elements using a kernel function, and then continuously morphs the faces between fluid elements over time such that the mass of each fluid element remains constant. Because the fluid elements have zero mass flux, this allows them to be tracked over time, thus making MFM a Lagrangian method. The Riemann problem is then solved across the element faces with a Godunov-type scheme (HLLC method). 

For the problem of gravitational collapse, Lagrangian codes are superior to Eulerian mesh codes because they conserve angular momentum which is crucial for accurately modelling this physical process. The Lagrangian predecessor of MFM is the Smoothed-Particle Hydrodynamics (SPH) method. The advantage of MFM over SPH is that MFM doesn't need an artificial viscosity term for numerical stability. This makes MFM a less diffusive method than SPH and better at preserving shock-fronts \citep{Deng_2019}. Preserving the sharpness of density gradients is another important requisite of accurately modelling gravitational collapse. In our model, we definitively need to avoid viscous effects that lead to unwanted redistribution of angular momentum. The specific differences in outcome between the MFM and SPH methods for our setup are briefly discussed in Appendix \ref{section:MFMvsSPH}. 

\subsection{Simulation Setup}

The initial pebble cloud is a uniform density sphere inside a Keplerian shearing box which approximates the tidal forces in the disk on the cloud, thus in rigid rotation at the Keplerian frequency. Though this initial condition is more idealistic 
than using pebble clouds formed by hydro simulations of disk instabilities, as done in \citet{Nesvorny2021}, it allows us to isolate the impact of the birth place of planetesimals on their properties. The rigid rotating pebble cloud in bound rotation around the star is not only the exact condition for the concept of hill density to trigger collapse \citep{Klahr_2020}, it also resembles the average pebble cloud and one can later study the effect of extra spin inside the cloud on the outcome of the collapse. 

All simulations discussed in this paper are done with $N=10^5$ particles, which is the maximum feasible resolution. The resolution convergence test is discussed in Appendix \ref{section:resolution}. With our model, this resolution is able to resolve tighter binaries than any previous simulations of planetesimal formation by gravitational collapse. We discuss this point further in Section \ref{section:binaries}. Note that our simulations are single-species, meaning we only model the pebbles and ignore the drag pressure on the pebbles from the gas in the disk during collapse. This follows \citep{Nesvorn__2010,Nesvorny2021} in which they found at high particle densities, drag forces become insignificant compared to collisions.
Furthermore, as argued in \citep{Klahr_2020,Klahr2021} the collapse only starts when the turbulent diffusion time of the pebble cloud is longer than its free-fall and thus during the collapse further feedback from the gas can be neglected. Therefore the mass of a gravitationally unstable pebble cloud depends on the turbulence of the gas, which in the case of SI corresponds to masses equivalent to 10 - 100 km sized solid objects.

Independent of mass, for a cloud in a circumstellar disk to collapse under its own self-gravity, the cloud's density must reach the threshold at which self-gravity overtakes the tidal forces in the disk. The initial density of our pebble cloud is set to this threshold. The traditionally used density threshold is the Hill density
\begin{equation}
    \rho_{H}=\frac{1}{\pi}\Big(\frac{3}{2}\Big)^2\frac{M}{a^3}
\end{equation}
where $M$ is the mass of the central star and $a$ is the semi-major axis of the cloud in the disk. The Hill density is derived from the Hill sphere, i.e.\ the sphere connecting the two Lagrangian points at either end of the central object's Roche lobe. A Roche lobe is the teardrop-shaped region around an object in a binary system formed by the system's line of equipotential, meaning anything in that region is gravitationally bound to that object. 

Note that by construction, an object's Hill sphere always extends beyond its Roche lobe. Hence, a particle can be inside an object’s Hill sphere yet outside its Roche lobe, resulting in that particle not actually being gravitationally bound to the object. Because of this, the Hill density is in fact an under-approximation of the critical density for gravitational collapse against tidal shearing. We use a more suitable estimation of the critical density using the sphere of equivalent volume to the object’s Roche lobe, the radius of which is derived in \citep{1971ARA&A...9..183P}. This "modified" Hill density is given by
\begin{equation}
    \rho_{H^*}=\frac{1}{\pi}\Big(\frac{3}{2}\Big)^5\frac{M}{a^3}
\end{equation}
The modified Hill density is a factor of $(3/2)^3$ greater than the traditional Hill density. Using this modified Hill density ensures that all particles in the cloud are gravitationally bound and will participate in the collapse.

\subsection{Polytrope Model}
It is not possible to treat all pebbles involved in the collapse as individual particles. Therefore one treats them rather as super-particles representing a cloudlet of pebbles. But one cannot replace such a cloudlet of the combined mass of its pebbles into a solid sphere of the same material density, as one would then chiefly underestimate the collision rates of the pebble cloudlets \citep{Nesvorny2020}.
And without collisions, the collapse would lead to a virialised cloud rather than a solid object \citep{2014Wahlberg-Jannson}. 

Previous N-body simulations of collapsing clouds consisted of collisional super-particles that represent sub-clouds of pebbles \citep{Nesvorn__2010,nesvorny2019transneptunian,Nesvorny2021}. During collisions, super-particles either exchange momentum based on a collisional time scale estimator or if particles get too close to each other, then they merge to create a new spherical super-particle with a mass based on the energy of the collision and previous masses. Instead of treating collisions directly, we opted for modelling our pebble sub-clouds as adiabatic gas particles. The pressure of an adiabatic gas is related to its density by the polytropic equation of state:
\begin{equation}\label{eq:eos}
    P=K\rho^{1+\frac{1}{n}}
\end{equation}
where $K$ is a constant related to entropy and $n$ is the polytropic index. 

A polytropic index of $n=0.5$ does for instance nicely describe the internal structure of the Earth \citep{OKeefe1966}. We use a slightly softer polytropic index of $n=1$ which still resembles the radial density profile of Earth (see Figure \ref{fig:polytrope} in Appendix \ref{section:EOS}), but reflects the compressibility of a granular object and leads to lower speed of sound, which impacts the computational time step. Our code solves gravitational equations simultaneously with the hydro equation (\ref{eq:eos}), allowing pressure-supported clumps to form from the initial sparse cloud when the gas reaches hydrostatic equilibrium (HSE). The constant $K$ can be set such that the gas stops collapsing when it reaches a density of $\rho_f=1.0$ g/cm$^3$, resulting in clumps with densities matching that of solid planetesimals. 

We normalized our calculations by setting the gravitational constant $G=1$, the central star mass $M_*=1$, and the orbital radius of the cloud $a=1$. This fixes the modified Hill density and therefore the initial density of our clouds, and it gives a ratio of the cloud's orbital period to free-fall time of $T_k/t_{ff}=18$. By also fixing the initial cloud radius, these simulation units allow us to use the same initial conditions for clouds with any central star mass and orbital radius. To differentiate clouds at varying orbital radii around a star of a given mass, we define the ratio of the initial cloud radius $R_0$ to the collapsed equilibrium radius $R_f$, defined for the case that all the pebbles would end up in a single body, as the \textit{collapse ratio} $\eta_c$
\begin{equation}
    \eta_c \coloneqq \frac{R_f}{R_0}
\end{equation}

In physical units, the final collapsed density of clumps is set to $\rho_f=1.0$ g/cm$^3$, and the initial density of the cloud is given by $\rho_{H^*}$ which depends only on the central star mass $M_*$ and the orbital radius of the cloud $a$. This means that from cloud to clump, the density will increase by a factor of $(\frac{\rho_{H^*}}{\rho_f})^{-1}$ and the radius will decrease by a factor of $(\frac{\rho_{H^*}}{\rho_f})^{(1/3)}$ (assuming that the density is constant over the volume). Therefore, where the cloud collapse is occurring in the parameter space of orbital radius and central star mass can be entirely described by the collapse ratio $\eta_c$.
 
The hydro-static equilibrium (HSE) density of clumps in our calculations is set by $K$, so to use $\eta_c$ as a variable parameter, a relation must be found between $\eta_c$ and $K$. Using the Lane-Emden equation to solve for the HSE value of $K$ for a clump with $\rho_f=1.0$ g/cm$^3$ gives the following equation for $K$:
\begin{equation}\label{eq:k}
    K=\frac{4\pi G}{(n+1)}\Big(\frac{\eta_c R_0}{\xi_1(n)}\Big)^2\Bigg[\Big(\frac{\rho_c}{\Bar{\rho}}(n)\Big)\rho_0\eta_c^{-3}\Bigg]^{1-\frac{1}{n}}
\end{equation}
where $G$ is the gravitational constant, $R_0$ is the initial cloud radius, $\rho_0$ is the initial cloud density, $\xi_1(n)$ is a dimensionless radius variable, and $\frac{\rho_c}{\Bar{\rho}}(n)$ is the ratio of the central to mean density of the final clump, for which we choose $\rho_c = \rho_f$. The terms $\xi_1(n)$ and $\frac{\rho_c}{\Bar{\rho}}(n)$ come from the Lane-Emden equation solution. The full derivation for Equation \ref{eq:k} is given in Appendix \ref{section:EOS}. 
 
We assume that all our initial pebble clouds are in bound rotation around the central star, which follows from the ansatz of Hill density and thus they will all have the same centrifugal radius $r_c$ to Hill sphere radius ratio.
The centrifugal radius for our pebble cloud is given as
\begin{equation}
    r_c \approx 0.13 R_0, 
\end{equation}
as we derive in the appendix. This means that if $\eta_c \ge 0.13$, then a single object can form from the collapse, which may be the case inside  $0.02$ AU. For large distances, $r_c$ will become much larger than $R_f$ and thus binaries will be born with larger separation and multiple fragmentation becomes more likely. This is the scientific question of this paper: The properties of binary and higher order planetesimal systems as a function of their distance to the sun. 

Ideally, for the correct pressure-density relation of solid planetesimals, the Tillotson equation of state should be used in combination with solving the equations for elasticity and plasticity of the body \citep{Schaefer2016}. However, this is beyond the scope of our current paper, in which we want to study the phase transition from collision dominated pebble cloud regime to a quasi-incompressible body. 

\subsection{Collisions}



The interaction between two mass elements in the GIZMO code is described via the acceleration according to the local pressure gradient
\begin{equation}\label{eq:p_collisions}
    \partial_t v = -\frac{1}{\rho} \nabla P = - K \rho^{\gamma - 2} \nabla \rho = - \frac{c^2}{\rho} \nabla \rho
\end{equation}
with pressure $P = K \rho^\gamma$, where $\gamma = 1+\frac{1}{n}$ and sound speed $c^2 = \frac{\partial P}{\partial \rho}$ 
. For $\rho = \rho_\mathrm{solid}$, i.e.\ when the body should be in hydrostatic equilibrium, one can show that $c$ is on the order of 
the escape velocity of the formed planetesimal. 
In the dilute phase, when $\rho < \rho_\mathrm{solid}$, $c^2 = \gamma K \rho^{\gamma - 1}$ is much lower than the typical relative velocities $v$ of the pebble clumps, as long as $\gamma > 1$. Thus shocks will occur, which mimic collisions in our model. 

Deviating from the continuity approach, we can consider Equation (\ref{eq:p_collisions}) as describing a repulsive force between two particles of mass $m$ and volume proportional to the cube of separation $r$ for a "subsonic" collision $c^2 > v^2$. We multiply both sides by $\rho$ and integrate over the Volume $V \propto r^3$ and get a formal equation for the acceleration of a super-particle with mass $m$ from the interaction with a second super-particle of the same mass.
 \begin{equation}
    m \partial_t v = - m c^2 \frac{1}{r}.
    \label{Eq:SPHkick}
\end{equation}

More rigorously, we could use the smooth particle hydrodynamics (SPH) approach of treating the pressure gradient in GIZMO 
in which the acceleration is proportional to pressure times the gradient of the kernel function $W$ \citep{Springel2005}, which also leads to Equation (\refeq{Eq:SPHkick}).
In a 2-body collision, such a repulsive potential of shape $c^2\ln{r}$ would perfectly model the elastic momentum transfer from one particle to the other and well approximate the collision of two hard spheres for sufficiently long time scales. Thus, a subsonic collision between two of our super-particles is already sufficiently modelled. With our polytropic equation of state which conserves entropy, we then find that subsonic collisions are perfectly elastic.

\citet{nesvorny2019transneptunian} emphasizes the importance of inelastic bounces among super-particles representing clouds of pebbles to efficiently cool their r.m.s.\ velocities which are powered by virialisation during collapse \citep{2014Wahlberg-Jannson}. At first glance, a polytropic equation of state implies perfect reversibility of any compression. Thus, there should be no damping of kinetic energy. Fortunately, the GIZMO code solves the Riemann problem for interacting particles, which only conserves entropy if the involved velocity difference is subsonic. For the densities at the start of collapse, the sound speed $c$ 
can be compared to a typical collision velocity, for which we choose the free-fall or escape velocity of a body with mass $m$ and density $\rho$. We calibrate the pressure, or more precisely $K$, for the formed planetesimal at size $R_f$ and solid density $\rho_f$  (see Equation \ref{eq:k}):
\begin{equation}
    K = \frac{2 G}{\pi}R_f^2 = \frac{3}{4 \pi^2 \rho_f} \frac{2 G \rho_f \frac{4 \pi^2 }{3} R_f^3}{R_f} = \frac{3}{4 \pi \rho_f} \frac{2 G m}{R_f} 
\end{equation}
Thus, we can express the entropy via the escape velocity     
\begin{equation}
K = \frac{3 v^2_\mathrm{esc}}{4 \pi^2 \rho_f},
\end{equation}
and directly express the speed of sound in terms of escape velocity
\begin{equation}
    c = \sqrt{\frac{3 \gamma}{4 \pi^2}} v_\mathrm{esc},
\end{equation}
Here we see that for the body at solid density $\rho_f$ the speed of sound is of order unity identical to the escape velocity.
For lower densities, i.e.\ at the start of the collapse, the speed of sound is given as
\begin{equation}
    c = \sqrt{\frac{3 \gamma}{4 \pi^2}} \sqrt{\frac{\rho}{\rho_f}} v_\mathrm{esc},
\end{equation}
which can be compared to the free-fall velocity during collapse:
\begin{equation}
    v_\mathrm{ff} = \left(\frac{\rho}{\rho_f}\right)^{1/6} v_\mathrm{esc},
\end{equation}
implying that the Mach number of a typical collision scales as:
\begin{equation}
    M_\mathrm{col} =\left(\frac{\rho}{\rho_f}\right)^{-1/3} = \eta_c^{-1}.
\end{equation}
So if our initial density is $\rho = 10^{-9}$, typical collisions are highly supersonic ($M = 1000$), and thus dissipate enough energy to form planetesimals, all while momentum, and more importantly angular momentum, is perfectly conserved. 
A collision with an impact parameter larger than zero is then an off-center flyby of dilute super-particles and happens typically at supersonic velocities $c^2 < v^2$. Here the Riemann solver automatically replaces the gas pressure by the ram pressure: 
\begin{equation}
    m \partial_t v = - m v^2 \frac{1}{r}.
\end{equation}
From that we can calculate the loss of kinetic energy in a collision as 
\begin{equation}
    \partial_t v^2 = - 2 v^2 \frac{v}{r},
\end{equation}
in which $t_\mathrm{coll} = \frac{r}{v}$ is a representative collisional time, giving a physical meaning to the decay of kinetic energy. 

In classical collision schemes \citep{Nesvorny2020} for super-particles (SPs), one starts with treating a system of real particles (RPs). Using a function of the RPs size distribution d$N$ and the cross section of the individual pebbles $\sigma'$, one can then calibrate a new cross section $\sigma$ for the SPs, which each contain $n$ RPs that receive the same loss of kinetic energy per time. \citet{Nesvorny2020} finds that $\sigma = \sigma' n$ is a good approximation for the case that all SPs have the same mass:
\begin{equation}
    \partial_t E' = - 2 m \frac{\sigma' n}{r^3} v^3 = - 2 m \frac{\sigma}{r^3} v^3,
\end{equation}
as is the case in our simulations.
Thus we can write our prescription as:
\begin{equation}
    \partial_t E' = - 2 m \frac{\sigma' n}{r^2} \frac{1}{r} v^3,
\end{equation}
to be identical to the classical collision scheme as long as $\frac{\sigma' n}{r^2}$ is of order unity.
We mimic the SP collision scheme without defining the actual number $n$ of RPs in our SPs, which would be given implicitly by their radius $a_0$ and the current cross section for the SPs $\sigma \propto a_0^2$ as $n = \frac{r^2}{a_0^2}$. So as soon as "solid" planetesimals of 100 km in diameter form and $r\approx a_0$, then we get $n=1$. 


For a proper treatment of the pebble cloud collapse, the ratio between collision time and free-fall time must be correctly captured.
The collision time for the initial pebble cloud at virial velocity is given as 
\begin{equation}
    \tau_\mathrm{coll} = \frac{4}{3} \frac{R_0^2}{n a_0^2} \Omega^{-1},
\end{equation}
which produces the same result for different pebble sizes $a_0$ as long as $n a_0^2$ is kept constant. 
We can also express the collision time as a function of $\eta_c$:
\begin{equation}
    \tau_\mathrm{coll} = \frac{4}{3} \frac{a_0}{R_0}\eta_c^{-2} \Omega^{-1},
\end{equation}
and we find that for a typical $\eta_c = 10^{-3}$, the entire expression 
$\frac{a_0}{R_0}\eta_c^{-2}$ equals unity for a cloud mass equivalent to a 100 km planetesimal consisting of 10 cm pebbles. The collision time is then
\begin{equation}
    \tau_\mathrm{coll} = \frac{4}{3} \Omega^{-1}.
\end{equation}
For our collision model using the shock treatment of the Riemann solver, we got $n a_0^2 = r^2$ which produces the correct collision time at the onset of collapse $r = R_0$:
\begin{equation}
    \tau_\mathrm{coll} = \frac{R_0}{r}\frac{4}{3} \Omega^{-1}.
\end{equation}
During the collapse the effective cross section in our model shrinks continuously to the physical cross section, when the final planetesimal forms.
This shrinking of $n$ can be regarded as coagulation during the collapse,
which otherwise is subject to modelling, because the cross section should not become larger than what is given by the square of SP separation.

The effective pebble size and cross sections during collapse scale as:
\begin{equation}
    \frac{3 n a^2_0}{r^3} = \frac{1}{r}
\end{equation}
and for $\eta_c = 1$ we have $n=1$ and $a_0 = r_0$ with $r_0$ the size of 1 element at solid density. That size depends on the number of particles $N$ we simulate and the compressed size of the pebble cloud as $r_0 = R_f N^{-\frac{1}{3}}$.
Also mass has to be conserved, thus $n a_0^3 = r_0^3$. We combine this into
\begin{equation}
    a_0 = R_f N^{-\frac{1}{3}} \eta_c^2.
\end{equation}
For $R_f = 50$ km and for $\eta_c = 10^{-2}$ that is $a_0 = 10$ cm.
And for $\eta_c = 10^{-3}$ that is $a_0 = 0.1$ cm. So at 30 AU at the onset of collapse the pebbles are a little too small, but during collapse they quickly "grow". 
Using too few particles would then lead to an even shorter collision time, and using too many particles, would underestimate collision, so one has to choose these numbers carefully. One can imagine compensating for this inside the Riemann solver, with a modification of the RAM pressure, but this shall be left to future work.



A more detailed collisional model would also have to include the size distribution of pebbles constituting the pebble cloud and even allow for pebble clouds of different mass. Also, the coefficient of restitution $C$ can not be meddled with in our set up. As we solve no explicit energy equation, we effectively have very inelastic collisions $C << 1$.
To summarize, our collision model neither defines a coefficient of restitution $C$ nor the typical pebble size $a_0$. Yet by construction it reproduces the correct collision time scale for the regime of interest, i.e.\ the collision time to be shorter than the free-fall time. If one would wish to simulate other cases, like much longer (or shorter) collision times, one would have to modify the Riemann solver, such that the RAM pressure would  be artificially lowered (or increased).
Regardless, this is beyond the scope of the present paper.

\begin{figure*}[hbt!]
\centering
    \includegraphics[width=\textwidth,keepaspectratio]{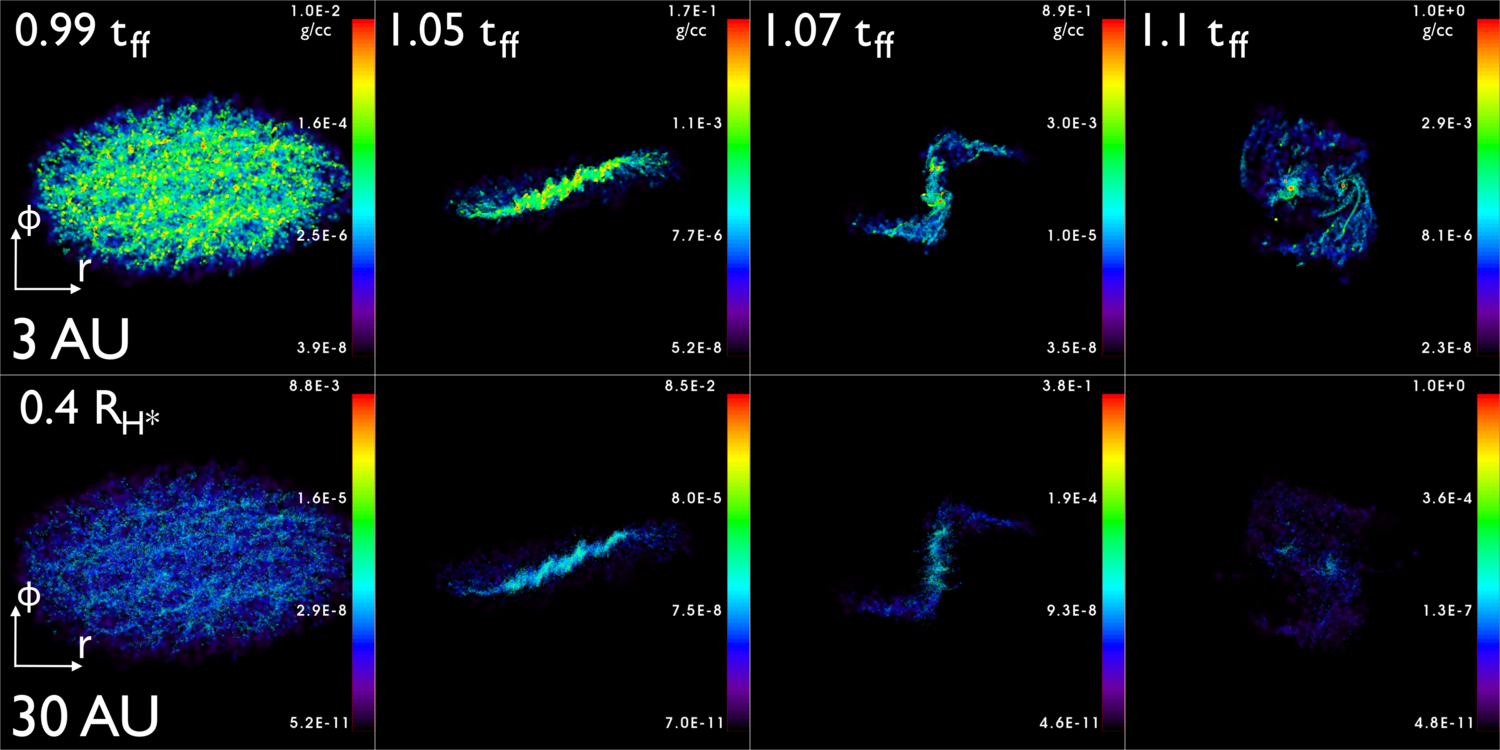}
    \caption{Density projection plots in the x-y plane of collapses at 3 and 30 AU for a 1M$_\odot$ star. For both the 3 and 30 AU case, times are given in the upper panel in units of free-fall time, and the scale of the horizontal radius is given in the bottom panel in units of the initial cloud radius. }
    \label{fig:etaTimes}
\end{figure*}


We performed tests of collapse simulations for various resolutions and found the dissipation of kinetic energy to be virtually independent for the range of resolutions we picked (see Figure \ref{fig:energyRes} in Appendix \ref{section:resolution}). This is a necessary condition for a super-particle collision model \citep{nesvorny2019transneptunian} which is met by our pressure and Riemann solver method for handling collisions. This approach intrinsically handles the collisions of pebble sub-clouds while also allowing the planetesimals to develop a physical size as they form, unlike the collisional super-particle method. 

Despite the adhoc ansatz of using pressure as a proxy for collisions, it has overwhelming benefits for the performance of numerical simulations. It treats collisions in a qualitatively correct way, dissipates kinetic energy lost by inelastic collisions, and it also provides a smooth transition from compressible to a quasi-incompressible planetesimal. The transition crosses up to 10 orders of magnitude in density in our simulations, and still the model provides a smooth adaptive resolution for the entire process.

\subsection{Clump Analysis}

To identify clumps, we use the \verb Scikit-learn  \citep{scikit-learn} implementation of the Density-Based Spatial Clustering of Applications with Noise (DBSCAN) algorithm \citep{10.5555/3001460.3001507}. DBSCAN is preferable to other clustering algorithms such as K-Means because it does not require the number of clusters to be specified. This allows the finder to be general enough to handle analyzing collapses that produce variable amounts of clumps. DBSCAN takes two input parameters, a distance and a minimum number of points, that act as a density threshold. The specifics of how DBSCAN works and how we use it are explained in detail in Appendix \ref{section:clumpfinder}. Before passing data to DBSCAN, we take an additional step and remove particles below a certain density threshold $\rho_{min}=0.1\%\eta_c^{-3}\rho_{H^*}$. 
Particles below this density are categorically not in clumps, and filtering them out improves both the efficiency and accuracy of the DBSCAN algorithm. In short, the clump finder routine we've built defines a clump as meeting the following three criteria:

\begin{enumerate}
    \item All points in a clump are $>=0.1\% \rho_f$
    \item Clump masses are $>=0.1\%\ M_{cloud}$
    \item All points in a clump are within a distance $R_f$ from at least one other clump point 
\end{enumerate}
In effect this defines the smallest clumps that we will identify to be $0.1 R_f$.
These clumps are still represented by up to 100 individual GIZMO particles, yet they will not contribute significantly to the cumulative mass distribution of objects, as we will show a posterior.

Once the individual clumps are identified, we determine the binaries and higher order systems in much the same way as \citet{Nesvorny2021}. We perform a simple double loop over all the clumps, starting with the most massive. A binary is confirmed if the two clumps have a negative binding energy and are within a Hill radius of each other. Then, an additional loop is performed over the binaries to find satellites of binary systems.

\subsection{Simulation Duration}

After the initial collapse phase when the particles become very dense, the evolution of the formed clumps  become significantly more computationally expensive. The time step is limited by the sound crossing time in our formed planetesimals, which is independent of the heliocentric distance $a$ of the particular system, yet the system evolves on the free fall and dynamical timescales at the given distance to the sun, which both scale with $t_\mathrm{dyn} \propto a^{1.5}$. Therefore a simulation at 30 AU is more than 460 times more costly than one at 0.5 AU. In order to compare our results we stopped all 
simulation at 1.43 free-fall times. At this time between 80 and 95 $\%$ of the pebbles are typically accreted into planetesimals (See Figure \ref{fig:etastats}). Ending the study here is justifiable in the sense that the scope of this study is only meant to characterize the properties of the "initial" planetesimal population that forms from the gravitational collapse of the pebble cloud. Of course one would also be interested in the longer term dynamical evolution of the planetesimal clusters, but this has to be done by other numerical means, like with a sink cell approach. Also, after a few free-fall times, the planetesimal cluster can no longer be considered as isolated from the protoplanetary disk environment, for instance other planetesimals.
Therefore, to study the long term evolution of the planetesimals, the external disk environment must be modelled and accounted for in dedicated N-body simulations. Pushing our simulations along would only result in a further dispersal of the initial population and perhaps some mergers. The future goals of this work are to take the initial planetesimal populations we formed in these simulations and insert them back into a disk environment to more accurately study their evolution and survival rate. 

\section{Results}
\label{section:results}

Here we discuss the results of our study. The complete list of collapse simulations and their properties can be seen in Table \ref{table:params}. Our simluations consists of two suites: 10 varying the location of the cloud in the disk (or the collapse ratio $\eta_c$), and 12 varying the random seed used for generating the initial particle positions in the uniform density spherical cloud. 

Each of these suites are analyzed and discussed in Sections \ref{section:birthplace} and \ref{section:stochasticity}, respectively. Section \ref{section:binaries} covers the properties of binary and higher order systems formed in all 21 simulations. Lastly, Section \ref{section:exo-p} describes how to apply and interpret these results for disks other than our solar system.

\begin{deluxetable}{l|ccc}
\tablenum{1}
\tablecaption{Simulations}
\tablewidth{\columnwidth}
\tablehead{
\colhead{Run} & \colhead{$\mathrm{log}_{10}(\eta_c)$} & \colhead{$a$ [AU]} & \colhead{seed}
}
\startdata
a0.5s0 & $-1.647$ & $0.5$ & 0\\ 
a1.0s(0-11) & $-1.948$ & $1.0$ & 0-11\\ 
a2.0s0 & $-2.249$ & $2.0$ & 0\\ 
a3.0s0 & $-2.425$ & $3.0$ & 0\\ 
a4.0s0 & $-2.550$ & $4.0$ & 0\\ 
a5.0s0 & $-2.647$ & $5.0$ & 0\\ 
a10.0s0 & $-2.948$ & $10.0$ & 0\\ 
a15.0s0 & $-3.124$ & $15.0$ & 0\\
a20.0s0 & $-3.249$ & $20.0$ & 0\\ 
a30.0s0 & $-3.425$ & $30.0$ & 0\\ 
\enddata
\tablecomments{$\eta_c=R_f/R_0$ is the collapse ratio of the cloud; $a$ is the orbital radius of the cloud for a 1$M_\odot$ central star; seed is the random seed used to generate the initial particle positions. A parentheses in the Run column indicates multiple simulations with the listed parameter varied.}
\label{table:params}
\end{deluxetable}

\subsection{The Impact of Planetesimal Birthplace}
\label{section:birthplace}

We simulated cloud collapses across the solar system, from 0.5 AU to 30 AU, in order to determine how the birthplace of planetesimals via the collapse ratio $\eta_c$ affects their properties and multiplicity. The orbital radius proved to have a profound influence on the dynamics of the collapse and the subsequent properties of the formed planetesimals. Figure \ref{fig:etaTimes} provides a general look at the dynamics of a cloud collapse in a circumstellar disk. Beginning with a uniform sphere, the cloud begins to collapse into an ellipsoid shape due to the tidal forces squishing the cloud in the z and radial directions. As the ellipsoid continues collapsing, first a disk and then a high density bar forms. Qualitatively, these are the same stages of collapse that were also reported on in the work by \citep{Nesvorny2021}. The tidal forces rotate the bar into a spiral, and several high density sub-regions within the bar collide along with premature clumps. The spiral does one more quarter rotation, which marks the end of the dynamical collapse. Clumps that have formed are left to interact with the other clumps and accrete the residual pebbles from the collapse on secular time scales.

\begin{figure*}[htb!]
\centering
    \includegraphics[height=0.8\textheight]{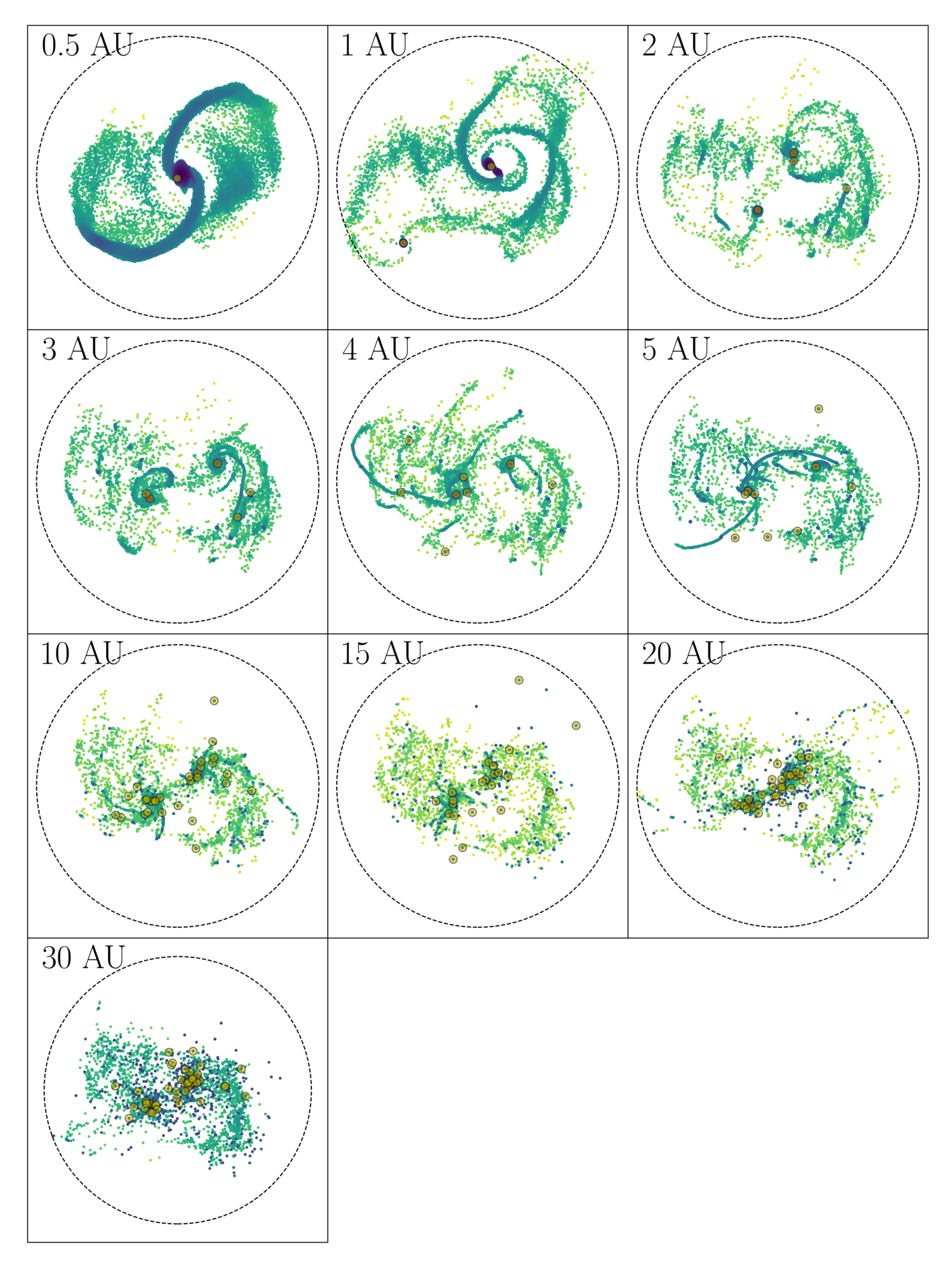}
    \caption{Views onto the plane of the disk of collapse simulations at various AU for a 1M$_\odot$ star. Color indicates density, the dashed black circles show the location and size of the initial cloud, and yellow circles are plotted over the formed planetesimals. The time of all the snapshots is when initial clumps form ($\sim 1.15 \mathrm{t_{ff}}$), which is when we begin tracking the clump properties. }
    \label{fig:etamorph}
\end{figure*}

Though the overall collapse proceeds similarly in both the 3AU and 30AU case, there is one key difference between them, as seen in Figure \ref{fig:etaTimes}. Throughout the entire collapse, there is a significantly higher degree of fragmentation in the 30 AU cloud. This manifests as both an increase in the number of forming clumps and a decrease in their average size. Similarly, the lesser amount of fragmentation results in fewer but larger clumps in the 3 AU collapse. Figure \ref{fig:etamorph} shows the initial clump formation states of the entire suite of collapse simulations at different heliocentric distances, with formed planetesimals indicated by faint yellow circles. This figure demonstrates the progression of fragmentation with distance from the sun. The rest of this section analyzes the properties of the formed planetesimals as a function of disk birthplace. 

Once planetesimals form from a cloud collapse, they mark the starting point for embryo formation. Unless some of the forming planetesimals are large enough to go straight into the pebble accretion phase \response{ \citep{OrmelKlahr2010,2017AREPS..45..359J,2017ASSL..445..197O,2022arXiv220309759D}}, which would be typically several 100 km, they first have to continue to grow via mutual non-destructive collisions, in an oligarchic phase. In this phase the size distribution of planetesimals stays narrow in shape, but shifts slowly towards larger sizes. Eventually one reaches a runaway phase \citep{Kobayashi2016,MORBIDELLI2009558}, at which individual planetesimals quickly outgrow their peers.
Because of this, constraining their initial mass and size distribution is essential for studying planet formation. 
In \citet{Klahr_2020} we derived a typical cloud mass that can undergo a gravitational collapse and the present project we want to learn what the overall efficiency and final outcome in mass function of that collapse can be.

\begin{figure}
\centering
    \includegraphics[width=1.0\columnwidth]{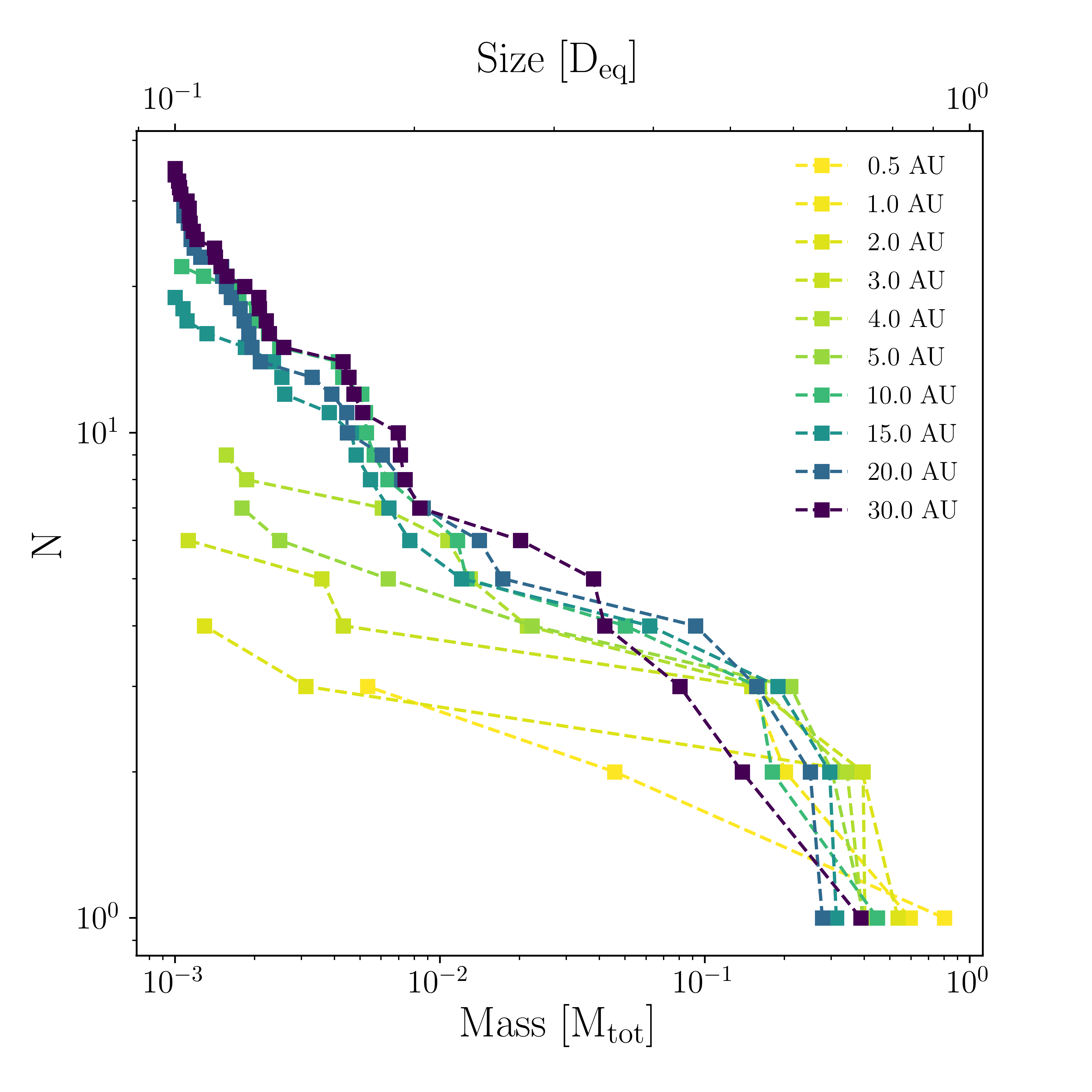}
    \caption{Size and mass distribution of clumps in units of the equivalent compresses size of the initial pebble cloud $D_{eq}$, respective pebble cloud mass $M_{tot}$ against cumulative number. Color represents orbital radius.}
    \label{fig:etaIMF}
\end{figure}

\begin{figure}
\centering
    \includegraphics[width=1.0\columnwidth]{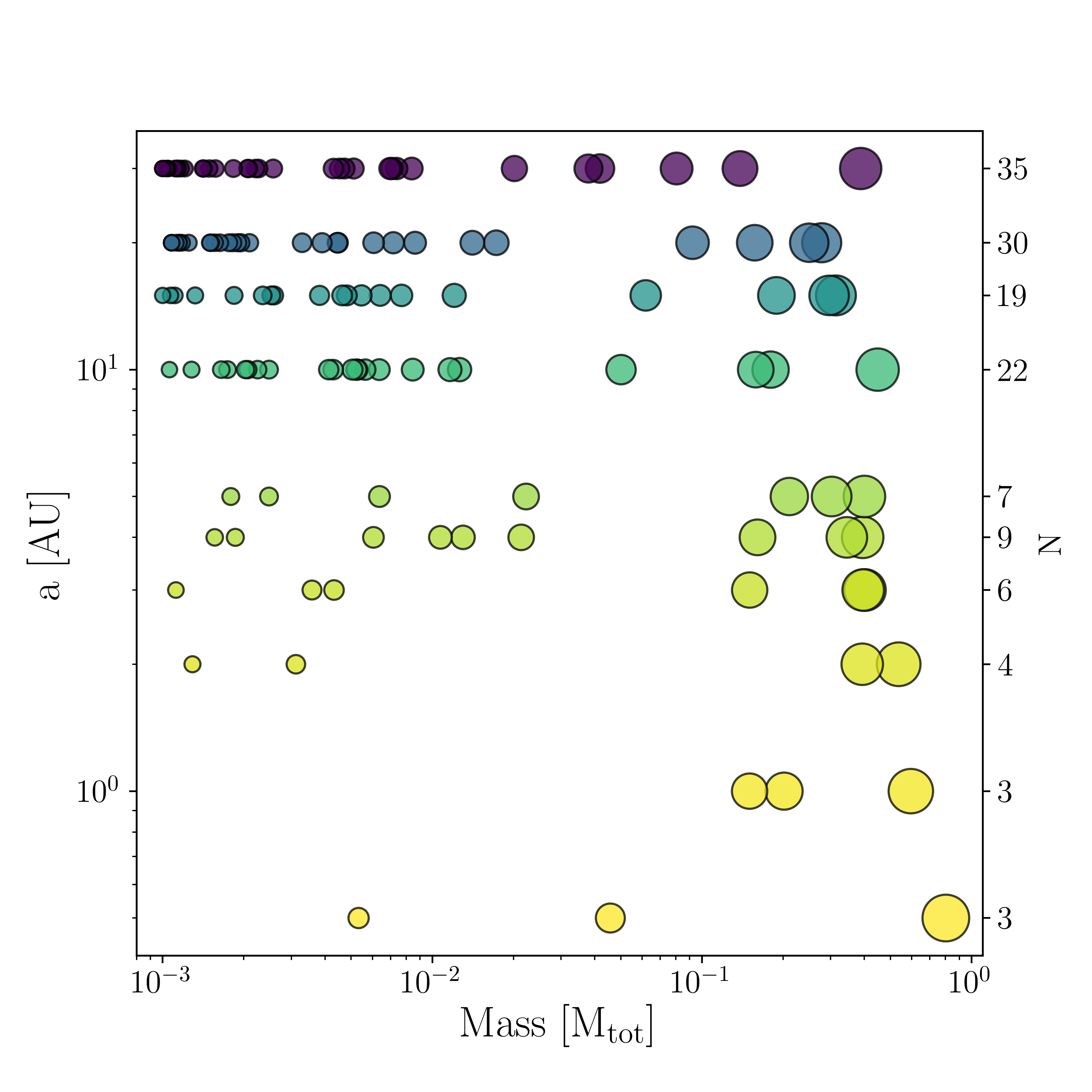}
    \caption{Visualization of the relative size and mass distribution of clumps. Color represents orbital radius.}
    \label{fig:etaSizeMass}
\end{figure}

The size and mass distribution of the planetesimals we formed and their dependence on orbital radius is depicted in Figures \ref{fig:etaIMF} and \ref{fig:etaSizeMass}. The size of each clump is determined by using its mass and applying a final uniform density of $\rho_f=1$ g/cm$^3$. 

\begin{figure}
\centering
    \includegraphics[width=1.0\columnwidth]{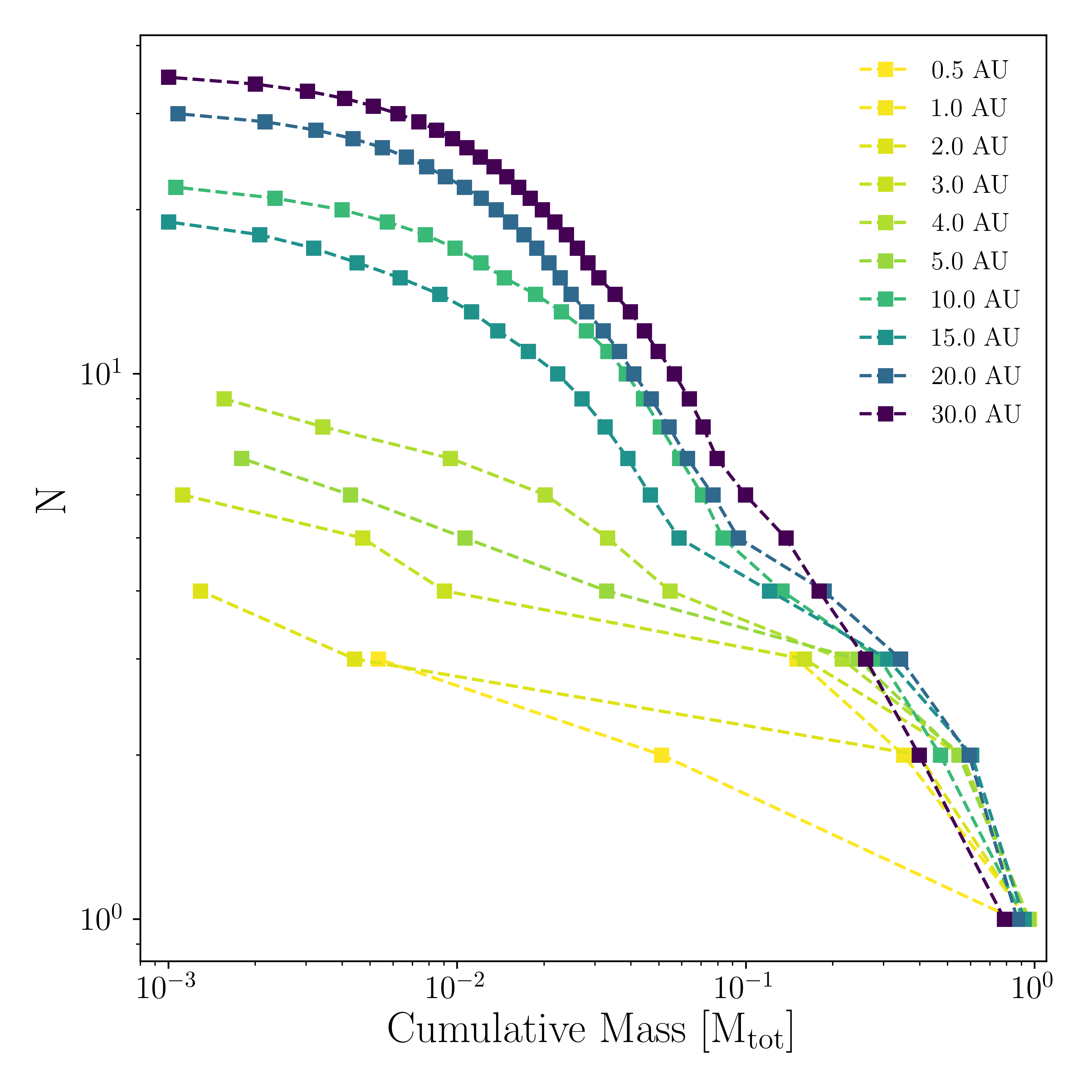}
    \caption{Cumulative mass distribution of clumps in units of the initial cloud mass. Color represents orbital radius.}
    \label{fig:etaCumMass}
\end{figure}

From Figure \ref{fig:etaIMF} we can see that regardless of disk birthplace, the largest $\approx 3$ planetesimals are always greater than 0.5 D$_\mathrm{eq}$, the equivalent diameter of the collapsed pebble cloud. Equivalently, from Figure \ref{fig:etaCumMass} we see that these same handful of large planetesimals contain a majority of the initial cloud mass. 

For many years, observations have shown that almost every minor solar system object family, from the asteroid to Kuiper belt, has the same characteristic size of $\approx 100$ km. Because these objects are planetesimal remnants, this implies that planetesimals have a preferred formation size regardless of their distance from the sun. Our results in combination with \citep{Klahr_2020} reproduce this phenomena, strongly suggesting that gravitational collapse is the predominant route to planetesimal formation.

This will then give us the size of the original pebble clouds that form the planetesimal population. \citet{MORBIDELLI2009558} found that most planetesimals are born big, with sizes in the range 100-1000 km. \response{This size range was further confirmed later by \citet{delbo2017}, and using those results we determined our equivalent diameter to be $\mathrm{D_{eq}}=200$ km.}

The size and mass of the planetesimals relative to the entire population is shown more clearly in Figure \ref{fig:etaSizeMass}. Bottom to top represents the growing distance from the sun. In this direction, generally there is a small decrease in the mass and size of the largest planetesimal. The largest planetesimals also tend to become more equal in size as the orbital radius increases. In all the simulations, there seems to be a gap between the large and small planetesimals occurring in the range $~10^{-2}-10^{-1}\ \mathrm{M_{tot}}$. There is also a steady increase in the number of small planetesimals ($<0.5\ \mathrm{D_{eq}}$) as orbital radius grows, with a significant difference for $<5$ AU than for $\ge 5$ AU.

\begin{figure}
\centering
    \includegraphics[width=1.0\columnwidth]{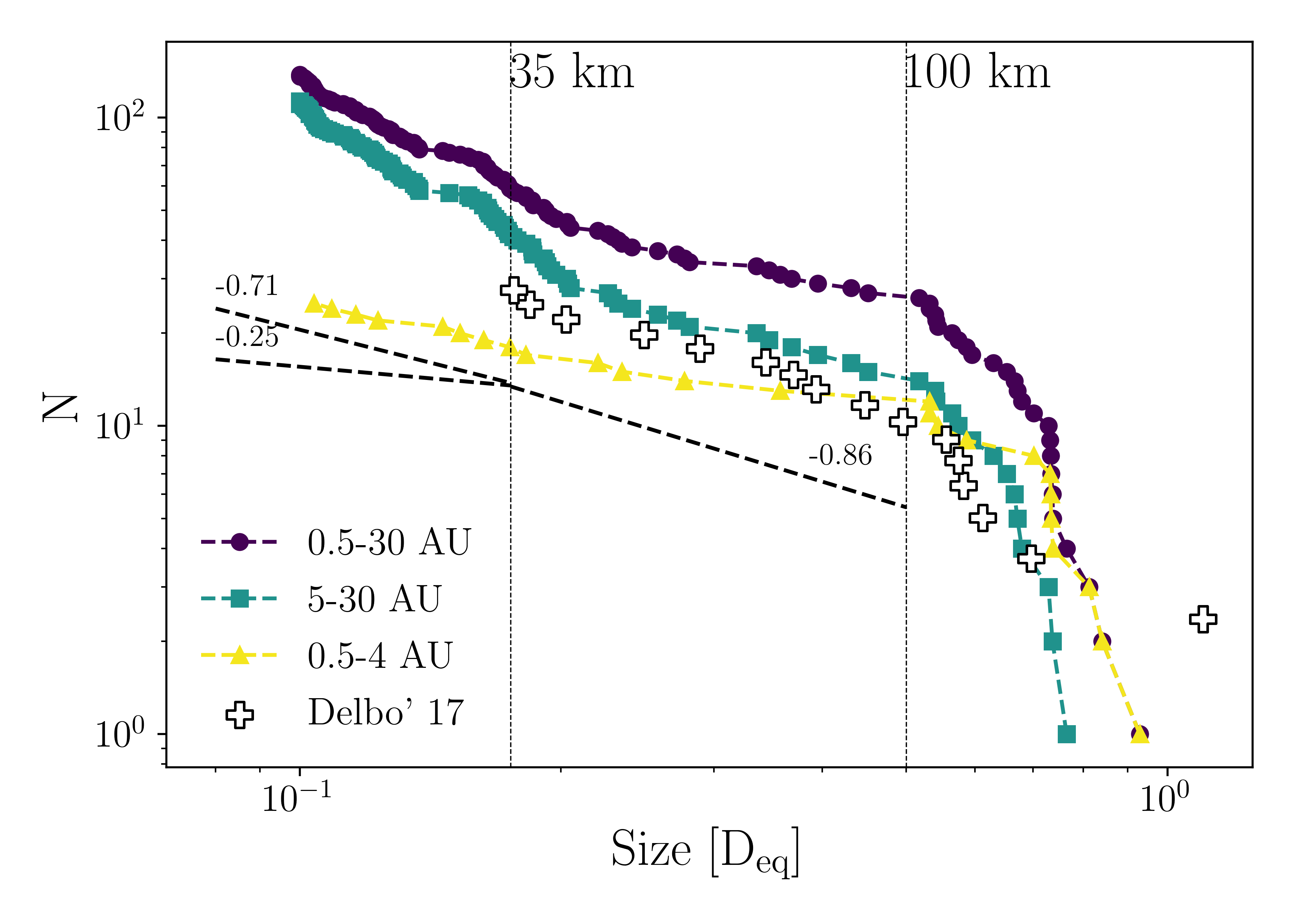}
    \caption{\response{Combined size distribution of all clumps formed in the simulation suite varying the orbital radius plotted with the size distribution of the original planetesimal population derived in \citet{delbo2017} shown by white crosses. Size is given in units of the equivalent compressed size of the initial pebble cloud $D_{eq}$. Color represents the region in the solar system the collapses occurred in. Simulations are separated into inner solar system ($<$5 AU), outer solar system ($\ge 5$ AU), and the entire solar system. The vertical dotted lines and power law dashed lines indicate the fits taken from \citet{delbo2017}. Our planetesimals follow the original population quite well. Planetesimals smaller than $0.5 D_{eq}$ have a shallower size distribution than $-0.86$, same as found for primordial asteroids. The smallest planetesimals, smaller than $0.75 D_{eq}$, fit the power law in the inner disk but the outer disk seems to produce a larger amount of small planetesimals than predicted.}}
    \label{fig:etaComIMF}
\end{figure}

This distinction for the size distribution for the inner and outer solar system can be seen more clearly in Figure \ref{fig:etaComIMF}. The objects formed in the five collapse simulations at 0.5-4 AU and the five at 5-30 AU respectively were collected into two groups and their size distribution was plotted. In both regions, the largest planetesimals are born big. In the inner disk, however, they are born slightly "bigger". In the inner region the largest planetesimals are $95\%$ of $\mathrm{D_{eq}}$, whereas in the outer part they are more like $75\%$ of $\mathrm{D_{eq}}$. Both regions have the same distribution shape for large objects, but the our region has a more pronounced tail in smaller planetesimals. 


\begin{figure}
\centering
    \includegraphics[height=0.8\textheight]{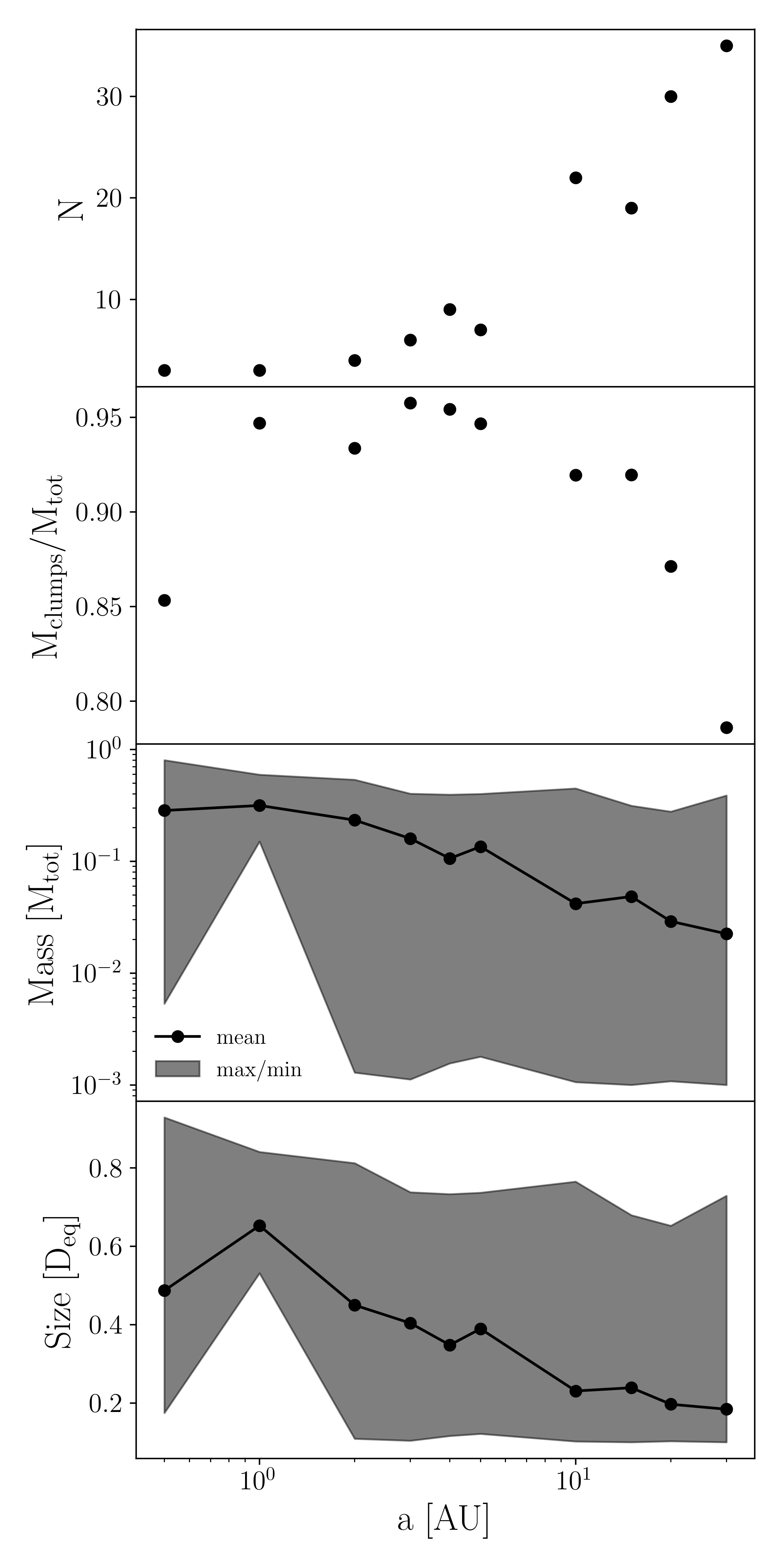}
    \caption{Characteristics of clumps formed against orbital radius. Properties shown are: (1) number of clumps, (2) total mass efficiency, (3) max., min., and average clump mass, (4) max., min., and average clump size.}
    \label{fig:etastats}
\end{figure}


\citet{delbo2017} identified a primordial family of asteroid fragments which is also very large and old. By eliminating this and other families of fragments in the same semi-major axis region they could identify a number of asteroids that they claimed cannot be included in any collisional family that had formed in locally in the main belt. Their logic consequence is that these asteroids that cannot be family members are some of the original planetesimals. From these asteroids the authors produced a size distribution of the original planetesimal population. In the size distribution of KBOs and asteroids there is a turning point at D$\approx$100 km, and in our simulations we see the same turning point at $\mathrm{D=0.5 D_{eq}}$. This implies that $\mathrm{D_{eq}\approx 200\ km}$.


Figure \ref{fig:etaCumMass} shows the cumulative mass of the planetesimals formed at different orbital radii in units of the initial cloud mass. There is a small but definite spread in the mass of the largest planetesimal, with the inner to outer disk corresponding to the largest and smallest. For all but the 0.5 AU case, the masses of the next largest planetesimal all follow roughly the same slope. Then, for distances below 5 AU the distribution flattens to lower masses, with a fairly large jump in the mass of the next clump. Above 5 AU, all the populations show the same mass distribution of clumps below $10^{-1}$M$_\mathrm{tot}$. Between 5-10 AU, there seems to be a distinct regime change in the initial mass function of planetesimals. This makes sense considering there is a significant difference in the characteristics of objects in the solar system that are above and below 5 AU.

Note that in our analysis, we include all formed planetesimals, bound and unbound. This means that if the simulations were continued, planetesimal mergers/accretion could change the overall distributions shown. However, to correctly constrain the effect of future mergers on the distribution evolution, the disk environment must be considered.

The clump properties analyzed by the previous plots in this section are compiled in Figure \ref{fig:etastats} where the following clump properties are plotted against orbital radius: number, total mass formation efficiency, and the maximum, minimum, and mean of the masses and sizes. This plot provides an overview of the trends we've identified for planetesimal formation properties depending on disk birthplace. One additional insight we can make from this plot is that for all orbital radii studied, the amount of mass from the initial cloud that ends up in clumps is $>75\%$. This confirms that gravitational collapse is a very efficient process in converting pebbles to planetesimals. 

\subsection{Stochasticity}
\label{section:stochasticity}

We analyzed a suite of 12 simulations that differ only in the random seed used to generate the initial positions of the particles in a uniform spherical distribution. This was done to measure the degree of stochasticity in a collapse and to see how much it affected the formation properties of planetesimals. The 12 simulations discussed in this section are at an orbital radius of 1 AU. This distance was picked for the statistical study because it is the computationally cheapest collapse that still produces multiple clump systems. 
\begin{figure*}[htb]
\centering
    \includegraphics[width=0.8\textwidth]{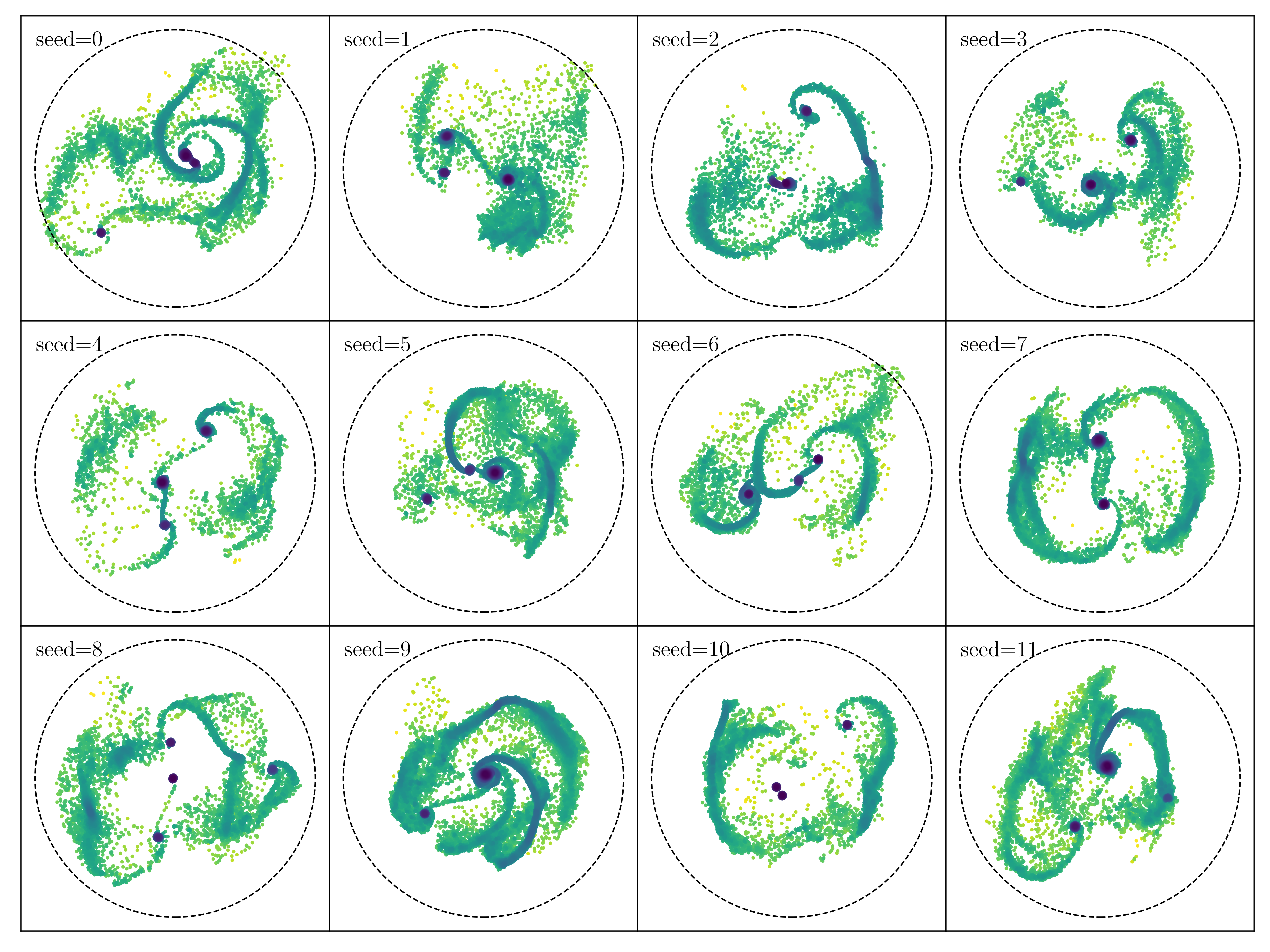}
    \caption{Views onto the plane of the disk of collapse simulations at 1 AU for a 1M$_\odot$ star. Color indicates density and the dashed black circles show the location and size of the initial cloud. The time of all the snapshots is when initial clumps form ($\sim 1.15 \mathrm{t_{ff}}$), which is when we begin tracking the clump properties.}
    \label{fig:seedMorph}
\end{figure*}
\begin{figure}
\centering
    \includegraphics[width=1.0\columnwidth]{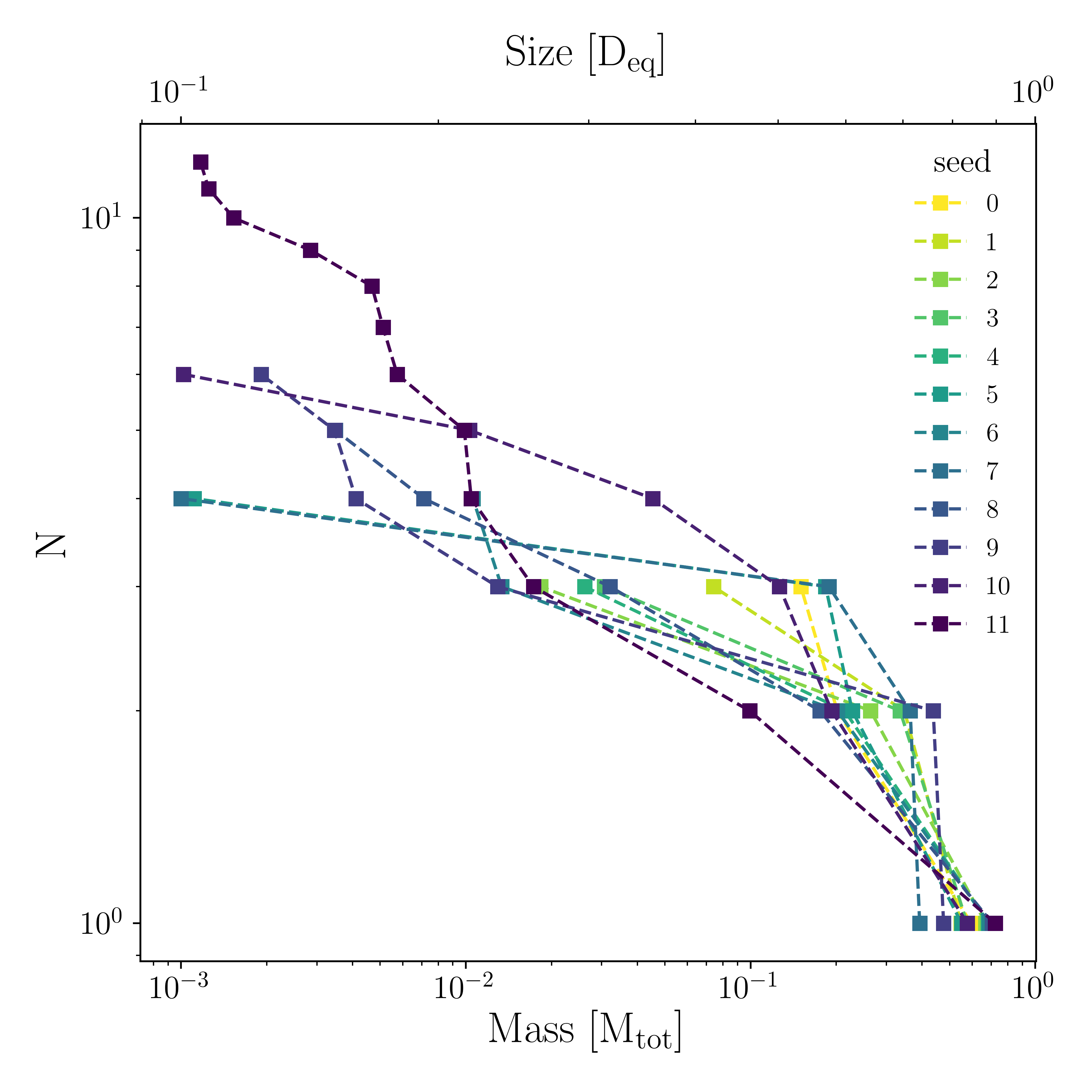}
    \caption{Size and mass distribution of clumps against cumulative number for simulations with different random seeds. Color represents seed.}
    \label{fig:seedIMF}
\end{figure}
\begin{figure}
\centering
    \includegraphics[width=1.0\columnwidth]{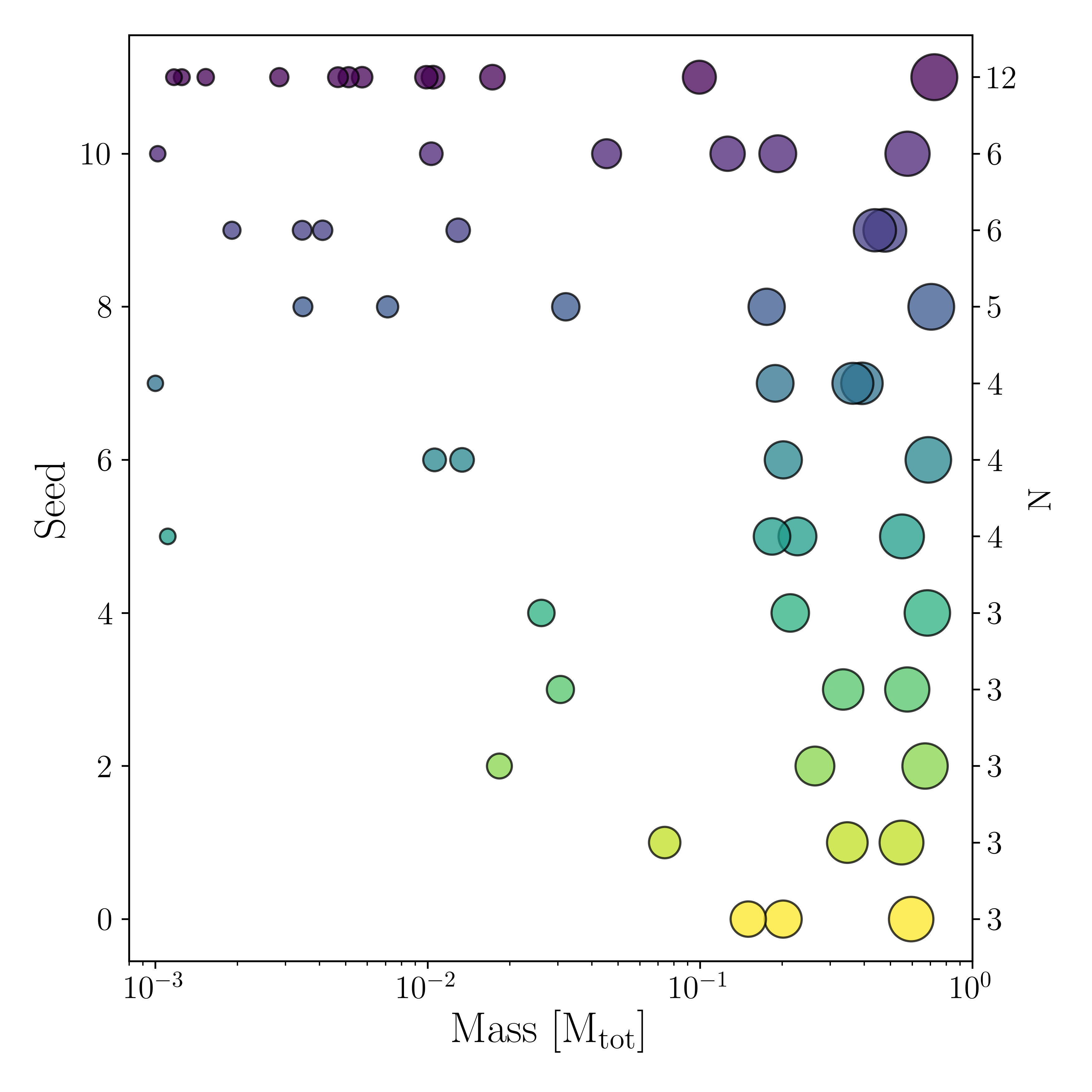}
    \caption{Visualization of the relative size and mass distribution of clumps formed in simulations with different random seeds. The color represents the seed value.}
    \label{fig:seedSizeMass}
\end{figure}
\begin{figure}
\centering
    \includegraphics[width=1.0\columnwidth]{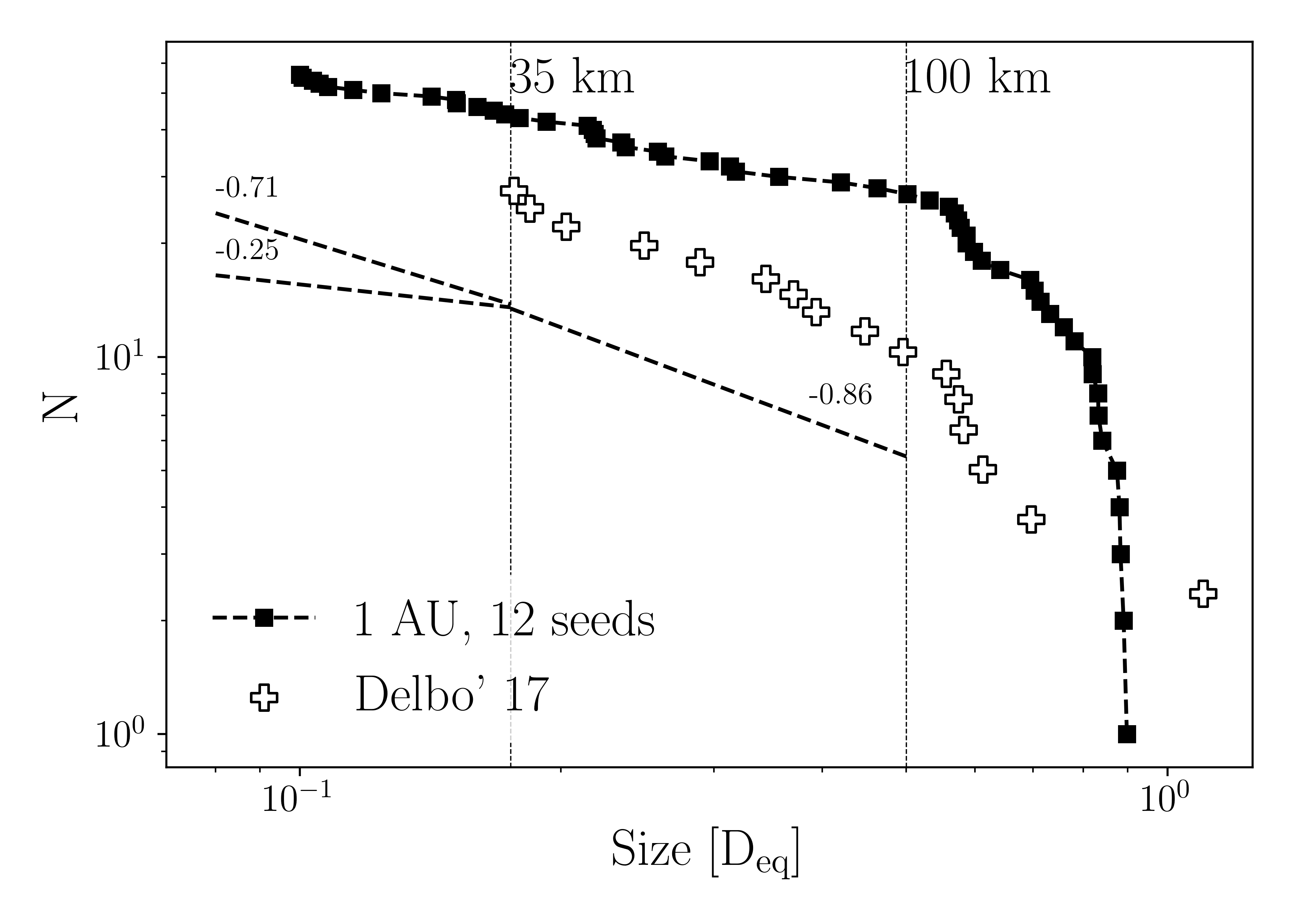}
    \caption{\response{Combined size distribution of all clumps formed in the 12 simulations with different random seeds along with the original planetesimal data from \citet{delbo2017}. The vertical dotted lines and power law dashed lines indicate the fits taken from \citet{delbo2017}. Planetesimals smaller than $0.5 D_{eq}$ have a shallower size distribution than $-0.86$, same as found for primoridial asteroids. Larger planetesimals follow $-0.86$ and steeper. The smallest planetesimals, below $0.175 D_{eq}$, lie within the predicted power law relations from \citet{delbo2017}.}}
    \label{fig:seedComIMF}
\end{figure}

\begin{figure}
\centering
    \includegraphics[width=1.0\columnwidth]{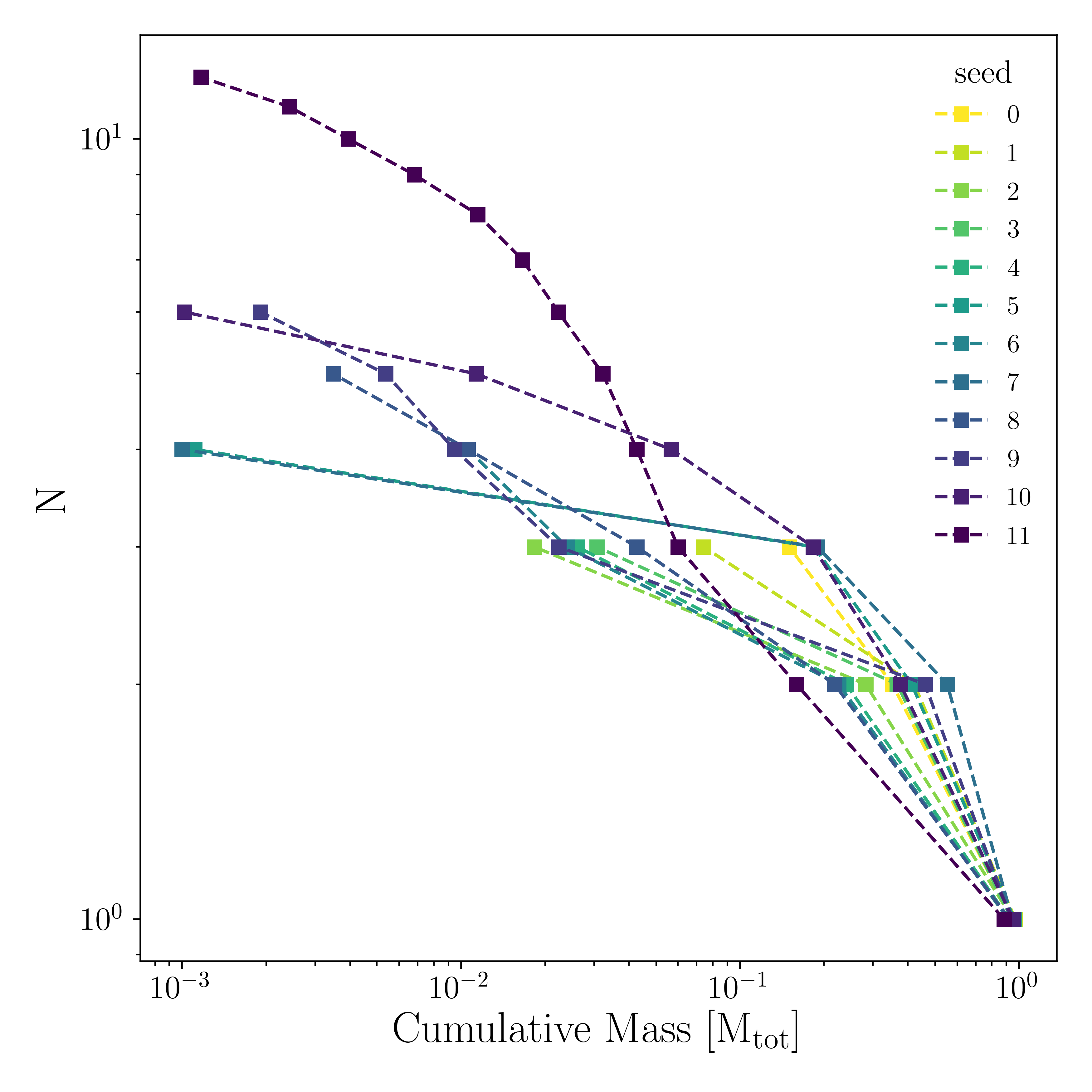}
    \caption{Cumulative mass distribution of clumps for 12 random seeds at 1 AU. Color indicates seed.}
    \label{fig:seedCumMass}
\end{figure}
\begin{figure}
\centering
    \includegraphics[height=0.8\textheight]{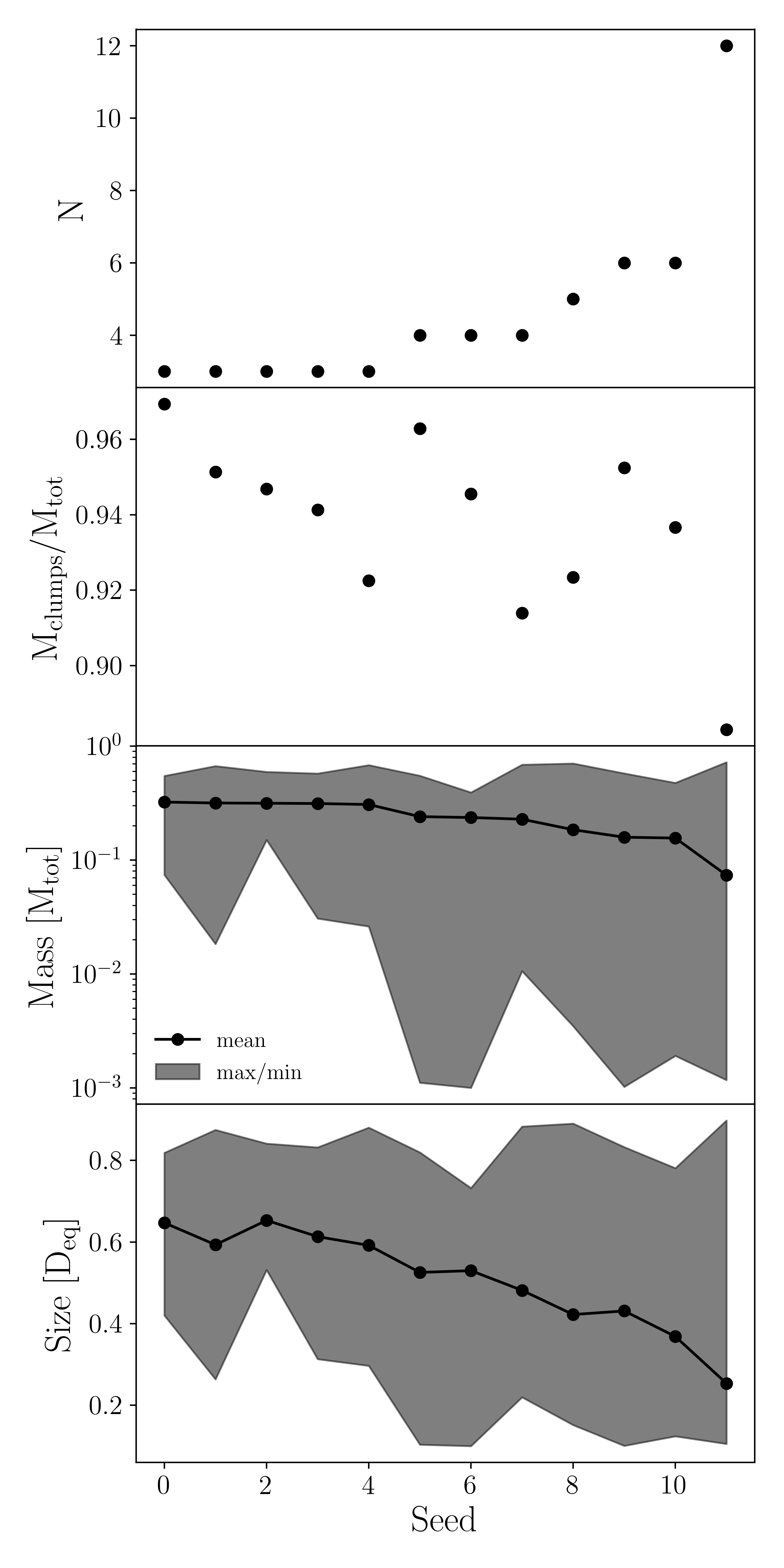}
    \caption{Characteristics of the clumps formed in each of the runs with different seeds.}
    \label{fig:seedStats}
\end{figure}

The analysis covered in this section matches exactly the one in Section \ref{section:birthplace}. This is done to serve as a benchmark for the consistency of the properties we saw in the previous set of plots. We can use the following results to determine which characteristics of the size and mass distribution of planetesimals formed across a disk are fundamental versus consequential of random perturbations in initial conditions.

Figure \ref{fig:seedMorph} gives a qualitative overview of the degree to which the random seed affects the final morphology of the collapse. Each of the snapshots are taken of the orbital (x-y) plane at 1.1 t$_{ff}$ when the clumps in each simulation first form. At first glance, though the orbital dynamics of the planetesimals vary (i.e.\ spatial distribution of the clumps), the number formed seems to be consistent at $3\pm1$.

Figures \ref{fig:seedIMF} and \ref{fig:seedSizeMass} show alternative views of the clump mass and size distribution in each of the 12 seed simulations. Most of the collapses produced 3-4 planetesimals (67\%), with a few forming 5-6 (25\%) and one forming 12 (8\%). All simulations produced 2-3 large planetesimals with sizes $\gtrapprox 0.5$ D$_\mathrm{eq}$. Though the number of clumps formed seems to be fairly stochastic, the mass of the largest clump in each of the 12 simulations is roughly consistent with an average value of $60\%\ \mathrm{M_{tot}}$ and a standard deviation of $9.6\%\ \mathrm{M_{tot}}$. The amount the largest clump mass deviates seems to depend on how comparable in mass the second largest planetesimal is. Because the mass of the largest clump is consistent for collapses with different seeds, the inverse relation we found between largest clump mass formation efficiency and orbital radius is valid even considering the stochastic nature of gravitational collapse. 

Figure \ref{fig:seedComIMF} shows the combined size distribution of all the clumps formed at 1 AU for the 12 different seeds, as well as the two size regimes and power law for the original planetesimal population of the asteroid belt derived in \citet{delbo2017}. The size regime above 0.5 D$_\mathrm{eq}$ again confirms the same steepness as the \citet{delbo2017} data, confirming this characteristic for collapse across the disk. The smaller planetesimals at 1 AU have a flatter distribution than the original planetesimal population, confirming the conclusion made from Figure \ref{fig:etaComIMF} that the slope of the small planetesimal ($<$0.5 D$_\mathrm{eq}$) size distribution steepens with increasing orbital radius. 

Figure \ref{fig:seedCumMass} shows the cumulative mass distribution (CMD) of the planetesimals. In comparison to the CMD for planetesimals at different orbital radii (see Figure \ref{fig:etaCumMass}), the same two regions of growth can be seen for masses above and below $~10^{-1}\ \mathrm{M_{tot}}$. For clumps above $10^{-1}\ \mathrm{M_{tot}}$, all seeds have the same shape with a slight spread in masses. For masses below $10^{-1}\ \mathrm{M_{tot}}$, there are 3 different patterns depending on how many clumps have formed. The CMDs with a smaller amount of clumps have a fairly steep slope with the smallest planetesimal being relatively large. Intermediate number CMDs have a flatter shape for the smallest planetesimals. And most notably, the CMD of the simulation that produces 12 clumps starts to resemble the same distinctive shape of the CMD seen for orbital distances at 10 AU and above. The fact that even at 1 AU, when enough small planetesimals form the CMD matches the CMD for $>10$ AU confirms suspicions of a universal initial mass function for planetesimals.

The compilation of clump properties formed at 1 AU with 12 different seeds is shown in Figure \ref{fig:seedStats}. The clump mass formation efficiency remains high regardless of seed, with all 12 simulations converting $\ge 88\%$ of the initial cloud mass into planetesimals. In addition to the mass of the largest planetesimal being roughly constant, the average planetesimal mass is also consistent despite the stochasticity of the collapse. \response{This spread in mass is seen more clearly in the average planetesimal sizes, where the average cloud will produce planetesimals of size 102 km on average with a standard deviation of 52 km. This is with $\mathrm{D_{eq}}=200$ km.} The average size of planetesimals decreases as the total number of them increases. This is consistent with the previous conclusion that a higher degree of fragmentation leading to more individual clumps forming will result in more mass being distributed to smaller planetesimals. This is the same trend we saw with the properties of clumps at different orbital radii, where the further out a collapse occurs, more planetesimals form that are on average smaller in size. 

To summarize, the stochastic nature of gravitational collapse does result in some variations in the mass distribution of planetesimals. However, the global properties of the planetesimal populations are ubiquitous despite small changes in the initial conditions of the cloud due to the use of different random seeds. This means that the overall trends and properties of planetesimal populations we identified in Section \ref{section:birthplace} are valid as inherent characteristics of collapses at the given orbital radius rather than random consequences of the initial setup. 

\begin{figure*}
\centering
    \includegraphics[width=1.0\textwidth]{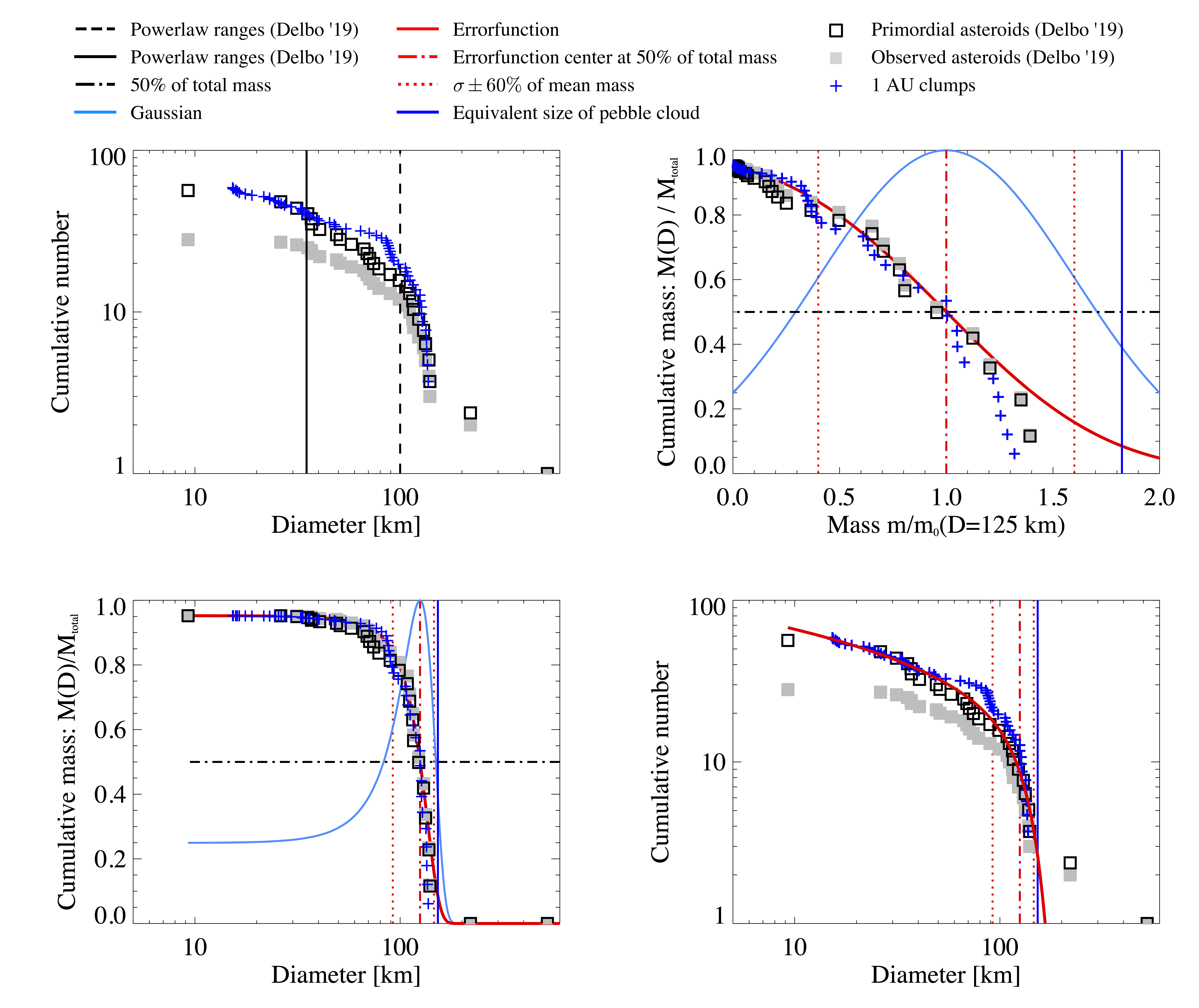}
    \caption{Combined size distribution of all clumps formed in the 12 simulations at 1 AU with different random seeds along with the "original" planetesimal data from \citet{Delbo2019}. Upper left plot: The cumulative number of the current and primordial asteroid population from \citet{Delbo2019} as grey (observed data) and black squares (number enhanced to accommodate for loss in catastrophic collisions), and the simulation data as blue crosses. Vertical lines are indicating the ranges where power laws were attempted to be fitted in \citep{Delbo2019}. Upper right plot: Cumulative mass of observational and numerical data compared with an errorfunction (red line), with its center at the $50\%$ level of total mass (red dashed dot line), equivalent to a diameter of $125 km$. Vesta and Flora were excluded from the cumulative mass, see text. The blue curve is the related Gaussian in total planetesimal mass per planetesimal mass bin. The vertical dotted lines indicate the standard deviation of $\pm 60 \%$ of the mean mass. The blue vertical line indicates the equivalent size of the original pebble cloud with $D = 152 km$. The lower left plot contains the same data, but plotted as function of asteroid size. And the lower right plot shows the cumulative number as in the first plot, but now with the overplotted analytical distribution as error function. A Gaussian in the mass distribution seems to describe the asteroid size distribution nicely as well as our simulation data for all asteroids smaller than the equivalent compressed size for the pebble clouds of $152 km$.}
    \label{fig:seedComIMFNEW}
\end{figure*}

\response{
\subsection{Initial size distribution of asteroids}
In the previous section, we did a qualitative comparison of our simulation data with the observational data in \citet{delbo2017} by defining a set of various power laws to characterize the distribution. This type of analysis is done not only on observational data, but also in the analysis of the mass distribution of self-gravitating pebble clumps generated in simulations of planetesimal formation \citep{Simon_2017,Liu2020}. In this section, we try a different approach to describe the size distribution of asteroids. We start from the idea that planetesimals form with a preferred mass, containing a certain fraction of the initial pebble cloud and that there is a Gaussian variation around this value.
For this purpose, we first define the cumulative asteroid mass per mass bin:
\begin{equation}
    M(D) = -\int_{Dmax}^{D} m(D) \mathrm{d}D.
\end{equation}
In the case that the mass distribution of asteroids is a Gaussian distribution of asteroid masses around median mass $M_0$ with a width of $\sigma_M$:
\begin{equation}
    M/dM = \frac{1}{\sqrt{2 \pi} \sigma_M} e^{-\frac{(M - M_0)^2}{2 \sigma_M^2}},
\end{equation}
then the corresponding cumulative function is the error function. 
Once the cumulative mass function for the asteroid data by \citet{Delbo2019} is calculated (see Fig.\ \ref{fig:seedComIMFNEW}), one can identify the most likely mass, $M_0$, where the cumulative function reaches $50 \%$ of the total mass in objects. 
In this fit, we removed the two largest objects in this set (Vesta with a diameter of $521 km$ and Flora with $220 km$\footnote{Todays diameter of Flora is about $146 km$, yet \citet{delbo2017} added some mass to estimate its original size, before it broke up in a collision creating the Flora family.}), which we suppose to be outliers of the distribution. This exclusion is mainly motivated by the fit of the Gaussian distribution, and thus is justified a posteriori.
In the case of the formation of Flora and Vesta, we argue that they were not primordial planetesimals, formed directly by pebble collapse, but that they formed by collisional evolution from smaller ($\approx 100 km$) sized precursors. In \citep{Voelkel2020a,Voelkel2020b}, we show how planetesimals grow from initial 100 km sizes via mutual collisions, before at some size, pebble accretion \citep{OrmelKlahr2010} takes over. Depending on where Vesta was formed and how big pebbles actually are, its size may still be smaller than expected for efficient pebble accretion. Furthermore, if Vesta would have reached a size large enough for pebble accretion while the pebble flux in the disks still lasted, it would have quickly grown to planetary embryo sizes. Furthermore, in the case of Flora, it is also possible it formed from the collapse of a higher mass pebble cloud.

Thus by excluding Vesta and Flora from the data set, we find the peak in the mass distribution at $M_0 = 125 km$.
The width $\sigma_M$ can likewise be found where the cumulative distribution reaches a value of $84 \% \approx 0.5 (erf{1/(\sqrt{2}\sigma_M)}+1)$.
Using the data from \citet{Delbo2019}, we find 
a width of $\sigma = 60\%$ of the mean mass (see Figure \ref{fig:seedComIMFNEW}). Overplotting the error function of the \citet{Delbo2019} data, we see a nice fit. From this, we see that in fact the primordial asteroids fall on a narrow mass distribution around 125 km sized objects.
When translating the error function in mass to the classical number per diameter plot (see \ref{fig:seedComIMFNEW}) we find also the smaller objects to be well described in the wings of the Gaussian.

Additionally, we included our 1 AU simulation data (blue plus signs) and find they also fit the mass Gaussian nicely. This allows us to derive an initial pebble cloud mass corresponding to a $152 km$ asteroid. Thus, the entire mass distribution of primordial asteroids can be explained by the stochasticity of the collapse from a single pebble cloud mass. Any additional masses of unstable pebble clouds as found in large scale simulations of the streaming instability \citep{Liu2020} would widen and alter the mass distribution of planetesimals, albeit to what degree would have to be constrained elsewhere.

So the $152 km$ equivalent size of the pebble cloud can explain all primordial asteroids, but Flora and Vesta. If they would also have been the product of a single pebble cloud collapse, the mass of this cloud would have to strong outliers and would be hard to explain why this massive clouds would not also have produced smaller companion asteroids showing up in the distribution.

By interpreting our numerical data from 12 collapse simulations forming 56 planetesimals we can interpolate that the 26 asteroids in the \citet{Delbo2019} data are the outcome of roughly 6 pebble cloud collapses all of similar size. In other words, each pebble cloud in the asteroid belt may have spawned on average 4 asteroids of the primordial population.
}


\subsection{Binaries}
\label{section:binaries}

\begin{figure*}[ht]
\centering
    \includegraphics[width=1.0\textwidth]{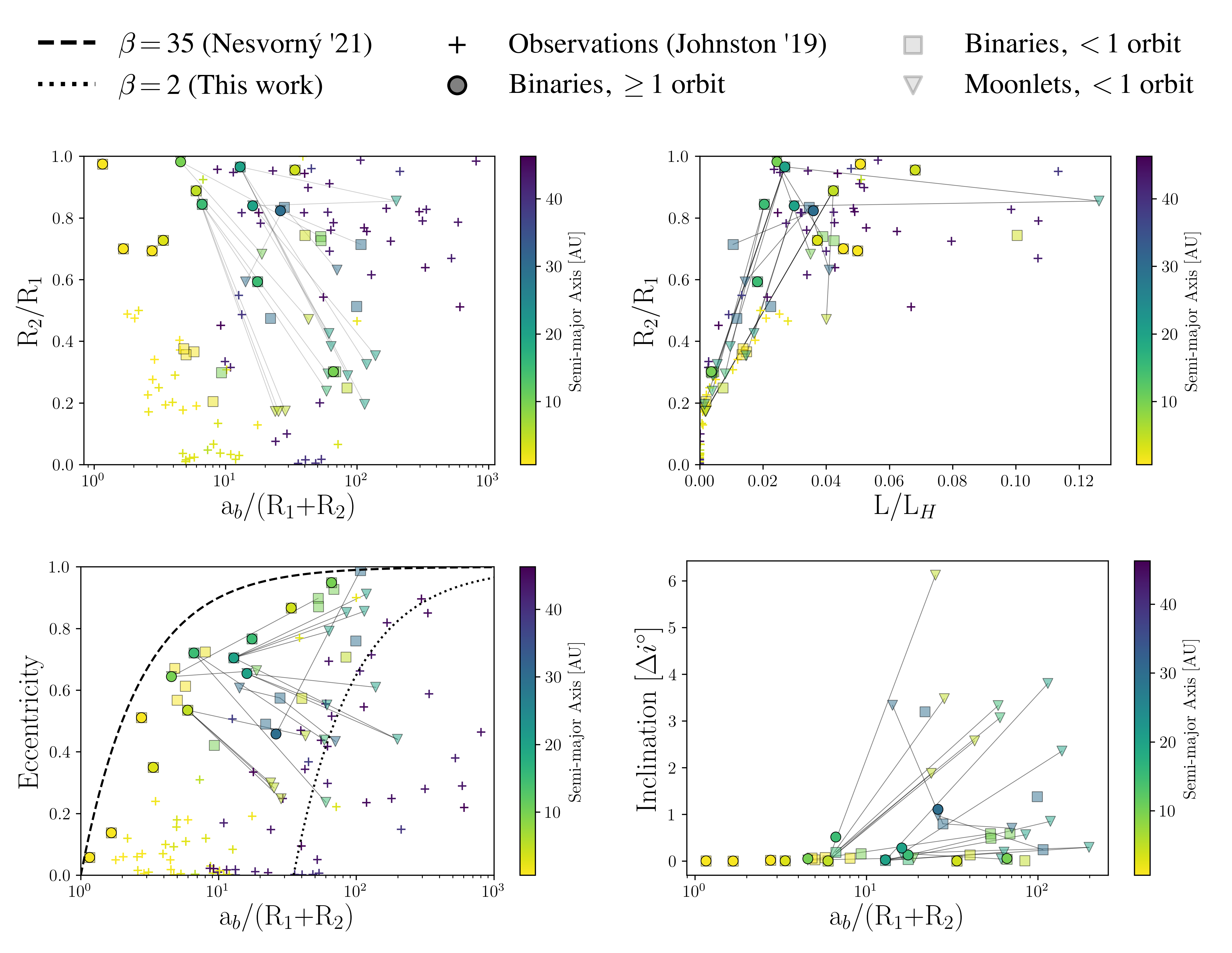}
    \caption{Properties of the binaries formed in our simulations. Circles, triangles, and squares are the simulated binaries and plus signs are observations of KBOs. The color represents the semi-major axis of the objects around a 1M$_\odot$ central star. The opaque circles are the binaries that have completed at least one orbital radius, and the transparent squares are the binaries that have been detected but haven't had time to complete an orbit. The transparent triangles are the moons, with the lines indicating to which binary they belong. None of the moons have completed an orbit. Upper left plot shows the scaled binary semi-major axis against the binary size ratio. Upper right shows the angular momentum scaled to the Hill angular momentum $L_H=$. Lower left shows the eccentricities. The lines denote the pericenter distances $\beta=a_b(1-e_b)/(R_1+R_2)=$35 (dotted) and 2 (dashed) which represent the binary separation resolving limit of the \citet{Nesvorny2021} simulations and our current model, respectively. This means our current model can resolve much tighter binaries than previous models. Lower right shows the binary orbit inclination angle from the initial rotation axis of the cloud.}
    \label{fig:binaryProperties}
\end{figure*}

So far we have characterized the properties of individual planetesimal clusters stemming from a single pebble cloud. In this section, we transition into discussing the properties of the binary and satellite planetesimal population. All of the binary analysis is compiled into Figure \ref{fig:binaryProperties}, in which the following properties are considered: size ratio, binary semi-major axis, angular momentum, eccentricity, and inclination. These plots were chosen specifically to compare our binary properties to those formed by the A12 simulations in \citet{Nesvorny2021}. Also shown by plus signs in Figure \ref{fig:binaryProperties} are the observed properties of known binary asteroids and KBOs, catalogued by \citet{2019pdss.data....4J}. The color of all the markers represents the heliocentric semi-major axis of the system in AU. 

\response{Of the binaries analyzed in Figure \ref{fig:binaryProperties}, the opaque circles are the tight binaries that formed with periods short enough to have their survival substantiated within the time frame of our analysis. Due to the high resolution of our study, the time span of our simulations is too short to be certain of the evolutionary survival of the wide binaries and satellites that have formed, shown by transparent squares and circles respectively. These wide binaries and satellites formed in collapses beyond 4 AU are distinguished in the plot because their surviving properties cannot be constrained within the timeframe of our simulations. This is because the long-term collisional survival of binaries in the outer solar system decreases with smaller object sizes and larger initial separations \citep{Nesvorny2021}. Therefore, the properties shown in Figure \ref{fig:binaryProperties} reflect tight binary properties and only the initial wide binary properties.}

The size ratio of binaries we produced favor equal size binaries with $R_2/R_1>0.5$ matching the observations of KBOs that also contain a majority of equal sized binaries. The outer disk binaries ($>4$ AU) match the tightest observed KBO binaries, occupying the same regions in the parameter space of scaled binary semi-major axis versus size ratio. \response{The observed outer disk KBOs with scaled semi-major axes around 50 do match the wide binaries we've formed. A few satellites even venture into the very wide region of the plot.}
The inner disk binaries on the other hand have roughly the same to slightly lower scaled semi-major axis and have a much higher size ratio than the observed binaries in the asteroid belt. This suggests that the survival rate of binaries in the asteroid belt is much lower than in the Kuiper Belt.
It seems that regardless of location in the disk, binaries favorably form with equal size ratios, and any deviation from that comes from its evolutionary lifetime in the disk.

In the angular momentum plot, our binaries only exist in regions where actual observed KBOs are. The simulated outer disk binaries follow the same relation as the observations where angular momentum increases with size ratio, as well as multiplicity increasing with both. While the outer disk binaries agree with the outer disk KBOs, the inner disk binaries have a much higher angular momentum than their counterparts in the asteroid belt. This again suggests that outer disk binaries don't have to undergo many evolutionary changes over the course of their lifetime, while inner disk binaries are more prone to property-altering processes. It appears that inner disk binaries lose angular momentum during disk evolution more efficiently, which is plausible due to many possible mechanisms like gas drag, radiation pressure or secular gravitational effects, all of which are more severe in the inner part of the solar nebula \citep{2021Icar..35613831L}.

On the eccentricity plot, there are two lines denoting the pericenter distances $\beta=a_b(1-e_b)/(R_1+R_2)=$35 (black, dotted) and 2 (black, dashed). These represent the binary separation resolving limits of our model ($\beta=2$) and the previous model ($\beta=35$) used by \citet{Nesvorny2021}. This means our models can detect and follow binaries that are $~18$ times closer than previously possible, and essentially reaching the maximum resolving limit because we can detect contact binaries. The only thing is that we need more resolution and a better pressure prescription for solids to track the \textit{evolution} of contact binaries. Nevertheless, we see a direct relation between increasing eccentricity and separation. As orbital radius increases, eccentricity increases along with binary separation. These same trends can be seen in the observations of KBOs. However, it seems that the eccentricity of close binaries decreases over the course of their lifetime in the solar system, and we do see this decrease in binary eccentricities begin to happen during the time frame of our study. The mechanism for this can be pebble accretion or interactions with satellites and/or other planetesimals passing by.

The inclination of every tight binary formed hardly deviates from the initial inclination of the collapsing cloud, with all deviating by less than $1^\circ$. Despite the relatively small range of inclinations, there is a clear trend that inclination increases with distance to the sun, related to the fact that more equally massed objects provide more degrees of freedom to distribute angular momentum. \response{We also see some satellites reaching inclinations as high as 6 degrees.} \citet{Nesvorny2021} studied the collapse of clouds with a distribution of spin orientations from SI simulations using an overall spread of inclinations over 360$^\circ$. Still the deviation of inclination between cloud and  binary was at most in the range of $\Delta i=10-20^\circ$, suggesting that either wider binaries tend to vary more in inclination, or that the initial spin of the pebble cloud has a profound impact on the forming binary, which we cannot test in the present paper.

\begin{figure}[ht]
\centering
    \includegraphics[width=1\columnwidth]{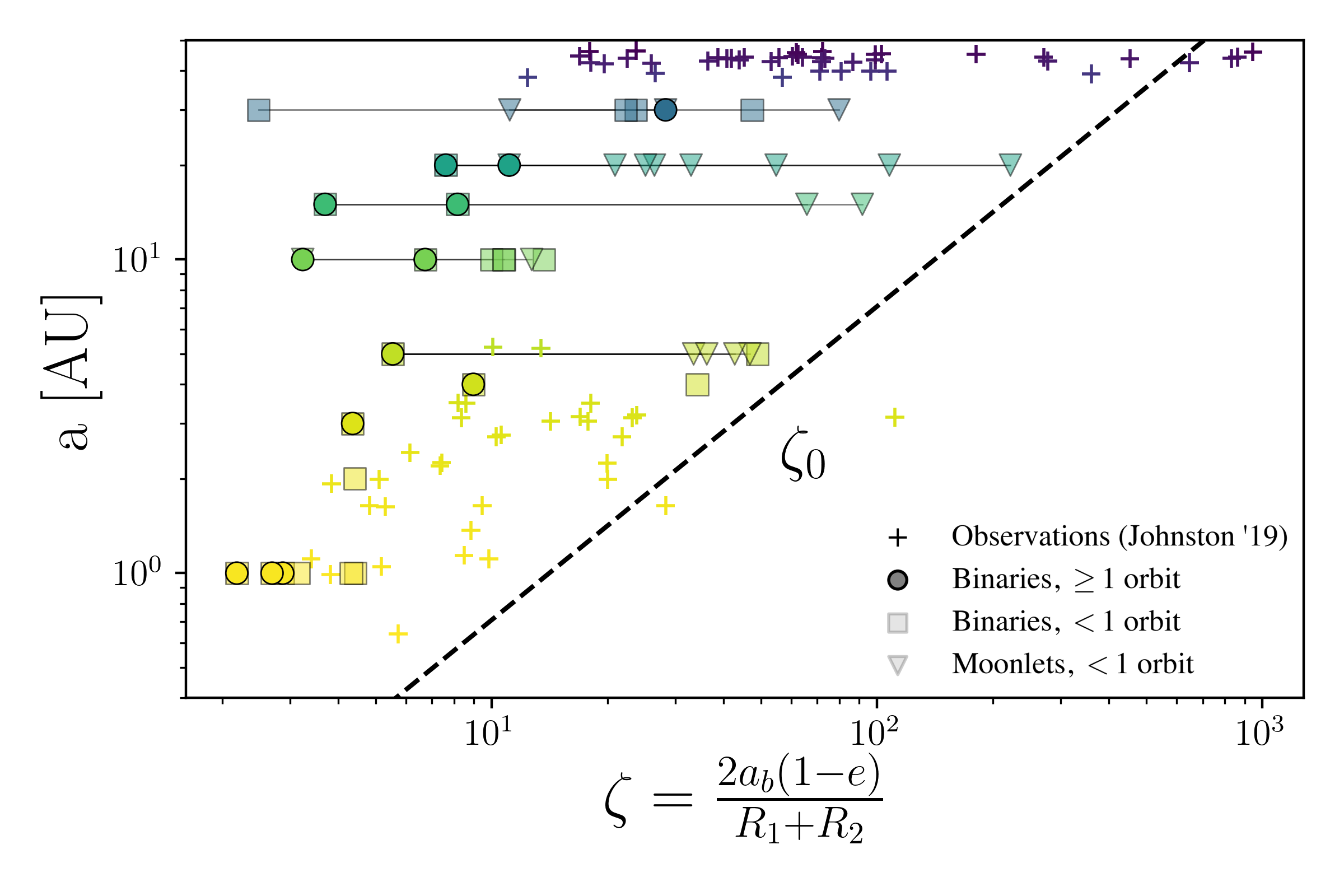}
    \caption{Binary proximity factor $\zeta$ against heliocentric semi-major axis. Circles, triangles, and squares are the simulated binaries and plus signs are observations of KBOs. The color represents the semi-major axis of the objects around a 1M$_\odot$ central star. The triangles are moons, with lines indicating to which binary they belong. The opaque circles have completed at lest one orbit, and the transparent squares (and triangles) have not. For a contact binary, $\zeta=1$. The maximum proximity factor $\zeta_0$ for a given semi-major axis is shown by the dashed black line. This factor corresponds to the binary that would form if all the angular momentum from the initial cloud went into a single equal mass binary.}
    \label{fig:binaryAU}
\end{figure}

An important aspect of planetesimal formation by gravitational collapse is the redistribution of angular momentum to many objects through multiple fragmentation. This dispersion of angular momentum is what allows the tight binaries we observe in our solar system to form. Figure \ref{fig:binaryAU} shows the proximity factor $\zeta=2a_b(1-e_b)/(R_1+R_2)$ which indicates the tightness of the binary, against the binary's distance from the central star. The minimum proximity factor is $\zeta=1$ which occurs for contact binaries. The maximum proximity factor $\zeta_0$ is dependent on the heliocentric semi-major axis and is denoted in Figure \ref{fig:binaryAU} by the dashed black line. This value represents the proximity factor for a binary that would form if all the angular momentum and mass from the initial cloud went into a single equal mass binary (see Appendix \ref{app:RCent}). From Figure \ref{fig:binaryAU}, all of the simulated binaries and almost all of the observed binaries have $\zeta<\zeta_0$. This suggests that the majority of KBO binaries originally formed alongside other binaries and planetesimals in a multiply fragmenting collapsing cloud. In both the simulated and observed binaries, the binaries tend to get wider as the maximum proximity factor $\zeta_0$ increases. However, most binaries are well below the $\zeta_0$ limit, and the further out binaries are relatively further away from the maximum proximity factor indicating a higher degree of fragmentation in clouds in the outer disk.


\subsection{Exo-Planetesimals}
\label{section:exo-p}

Because we fixed the initial modified Hill density of the cloud, our results can be applied to disks with different central star masses at a certain orbital radius. 

\begin{figure}[hbt]
\centering
    \includegraphics[width=1.0\columnwidth]{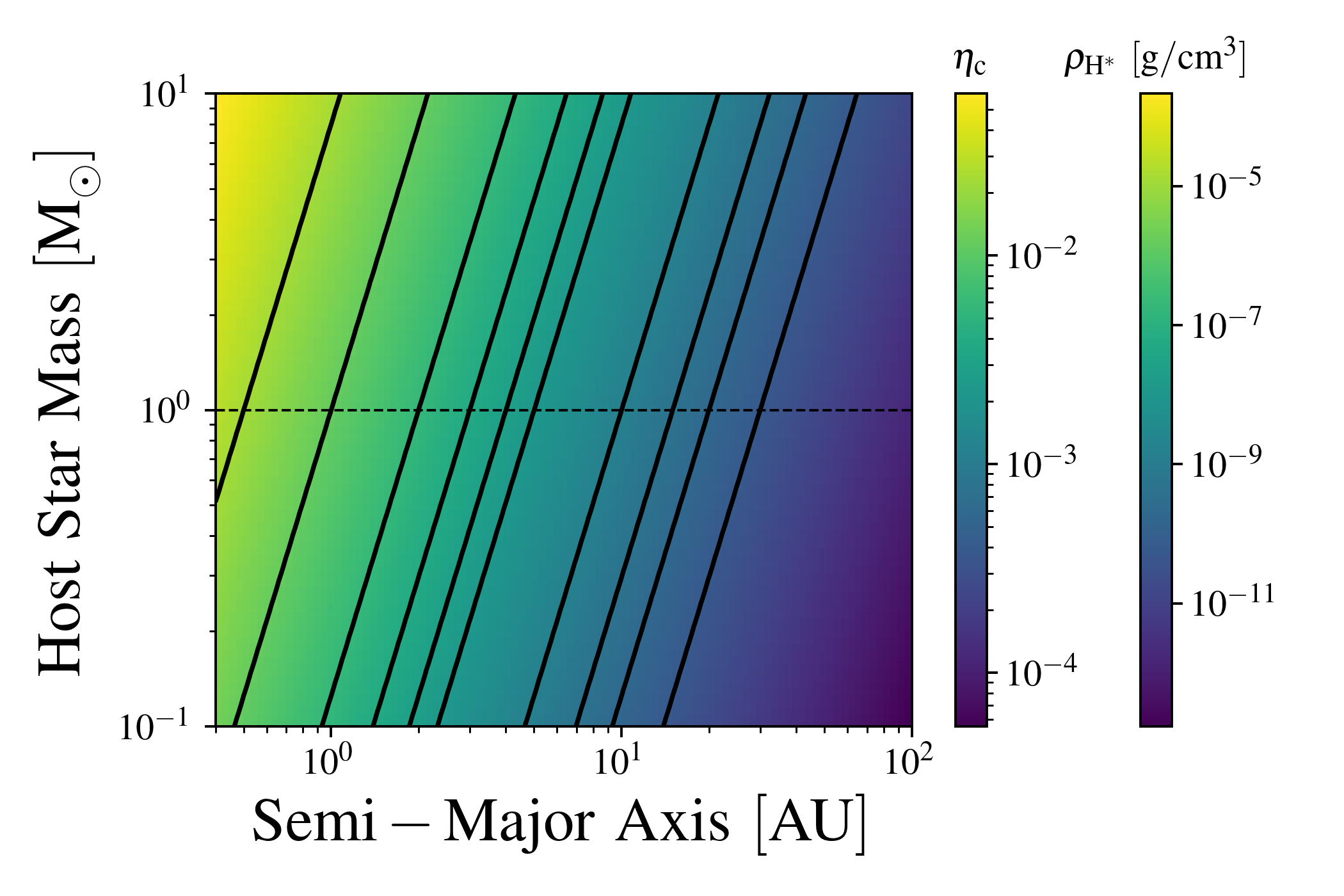}
    \caption{Physical values of the modified Hill density for any given central star mass and orbital radius. Black lines correspond to the equivalence lines of the $\eta_c$ values simulated in this study.}
    \label{fig:units}
\end{figure}

The conversion from the collapse ratio $\eta_c$ to a physical location in the disk can be seen in Figure \ref{fig:units}. Note that there are multiple modified Hill densities for a given $(M_*,a)$, meaning simulations with the same collapse ratio are applicable at multiple combinations of central star mass and orbital radius.

\begin{figure}[htb]
\centering
    \includegraphics[width=1.0\columnwidth]{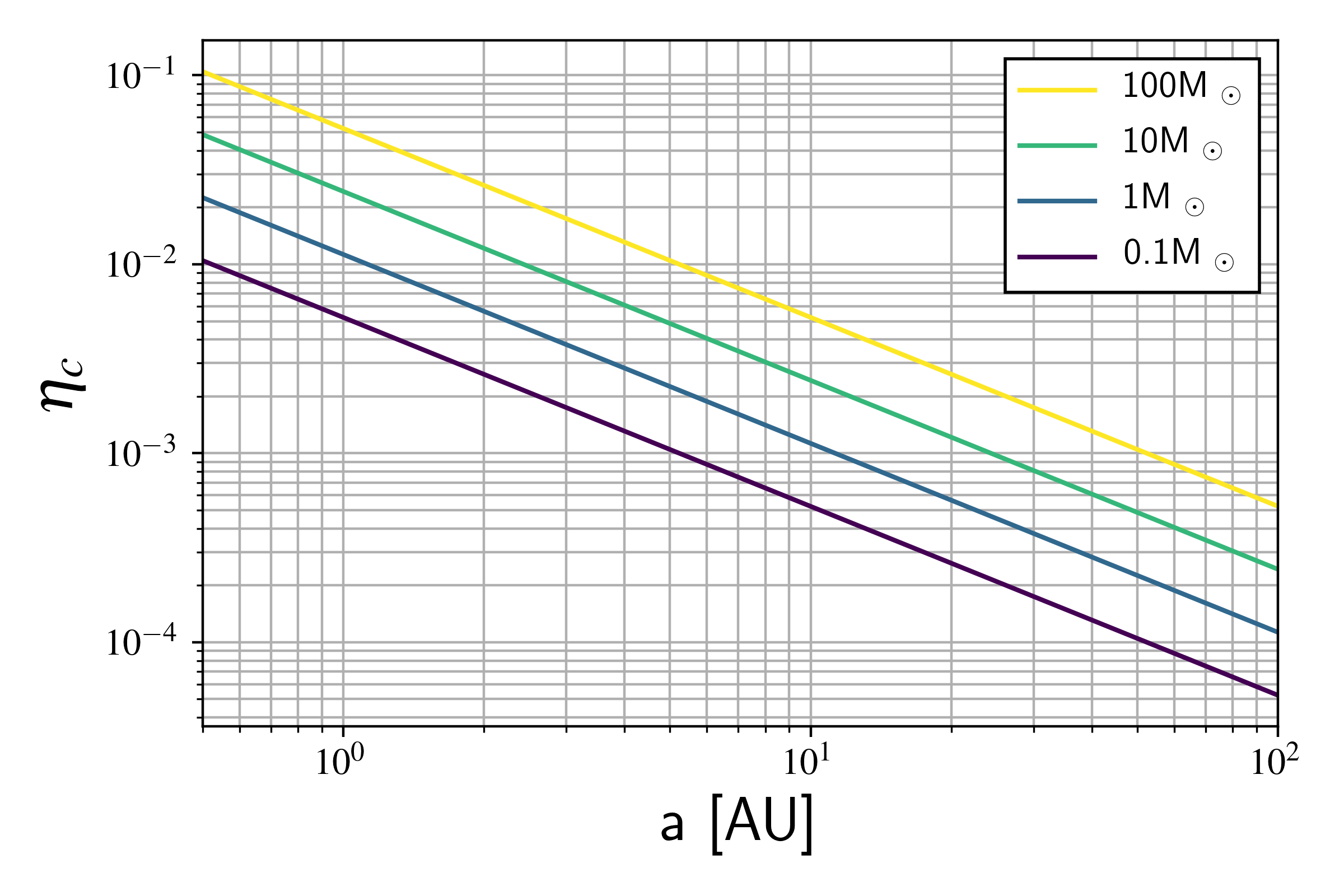}
    \caption{Plot showing the conversion from collapse ratio $\eta_c$ to orbital disk radius $a$ for various central star masses.}
    \label{fig:eta_a_rel}
\end{figure}

This means that to apply our model to exo-planetesimal formation in extrasolar disks, the only parameter to consider is the mass of the central star. To do this, we can use Figure \ref{fig:eta_a_rel} to convert our results for a 1 M$_\odot$ star to another star mass. For instance, consider our 30 AU cloud collapse around the Sun. This corresponds to an $\eta_c$ value of roughly $7\times 10^{-4}$. This means that a cloud collapse at 30 AU in  our solar system corresponds to a collapse at $\approx 15$ AU in a 0.1 M$_\odot$ disk and $\approx 60$ AU for a 10 M$_\odot$ disk. 

\section{Conclusions}
\label{section:conclusion}

In this paper, we have performed numerical simulations of planetesimal formation via gravitational collapse of pebble clouds. 
In a series of simulations, we always keep the same mass of the pebble cloud but vary the location of collapse, i.e.\ the distance to the young sun.
The heliocentric distance determines the initial size of a pebble cloud that can undergo collapse via the Hill criterion.

Thus, we can characterize the properties of the forming planetesimals depending on the location in the solar nebula. The overall conversion rate from pebbles into planetesimals is larger than $75\%$ for all distances. Also, the size of the largest planetesimals is in all cases close to the equivalent size of the pebble cloud, the size if the cloud was directly compressed into one single planetesimal.

The biggest influence the location of collapse has can be found in the multiplicity of planetesimals. Close to the sun, one is able to form a single object out of the collapsing cloud. But for large distances, an entire birth cluster of planetesimals, with a hierarchical sub-structure, is the typical outcome of collapse. Still, the mass distribution is always top heavy, with $80\%$ of the mass in the largest four objects. This means that the equivalent size of the pebble cloud remains a good estimator for initial planetesimal size \citep{Klahr_2020}. 

All of our simulations start with the same initial condition, given by the local Hill density and the global rotation of the nebula. However, the Hill density decreases with distance to the sun as $\propto a^{-3}$, and the dynamic range in density for the collapse increases. In other words, a collapse close to the sun leads to a solid object forming much earlier than collapses far away. In contrast to other simulations of pebble cloud collapse using sink cells, we were able to capture the moment of conversion from a compressible pebble cloud to a quasi-incompressible planetesimal.


The main results of our study can be summarized by the following points:
\begin{itemize}
    \item The distance of the cloud from the sun determines the number of planetesimals that will form a single cloud. Considering a cloud of a given mass, its collapse will produce fewer planetesimals at 1 AU and a greater number of planetesimals at 30 AU. 
    \item The size of the largest planetesimal scales with distance to the sun as $a^{-0.6}$. This would mean that for the same initial pebble cloud mass, the largest planetesimal at 10 AU may be 18$\%$ smaller than one forming at 1 AU. 
    \item Around 80\% of the total mass is contained within four or less planetesimals, regardless of the collapse location. This agrees with findings that despite huge distances between them, asteroids and KBOs generally have the same size, further suggesting that throughout the disk planetesimals are born big \citep{MORBIDELLI2009558}. 
    \item The planetesimal formation mass efficiency appears overall independent of location, with every collapse converting $>90\%$ of the initial cloud mass into planetesimals. For the planetesimal at $0.5$AU the efficiency appears as low as $85\%$, because with a dominant central body a circumplanetesimal pebble disk can form. Also for simulations beyond 20 AU the measured efficiency decreases down to $80 \%$ as the virialisation of the clusters is able to scatter more pebble cloudlets outside the region of gravitational influence of the planetesimal birth cluster, i.e.\ the Hill sphere.
    \item We produce many tight binaries, in some cases with a binary semi-major axis of $<1.5(R_1+R_2)$. In these cases we find the excess angular momentum of the pebble cloud to be stored in other components of the planetesimal cluster. Thus, there is no immanent need for an overly effective de-orbiting mechanism for wide binaries to explain the high frequency of bi-lobed objects and contact binaries among KBOs like Arrokoth.
    \item Many of the binaries we formed in the outer solar system have satellites. This falls along the findings of \citet{Nesvorny2021} and their prediction of the presence of such satellites in the Kuiper Belt.
    \item Though gravitational collapse is an inherently stochastic process, the general size distribution of the largest planetesimals remains the same for a collapse at a given orbital radius regardless of random seed. There also appears to be a preferred range for the number of clumps produced, most likely related to the fixed mass efficiency for the largest planetesimal.
    \item The properties of binaries produced by our model agree with observations of the binary properties of Kuiper Belt Objects \citep{2019pdss.data....4J}. 
    \response{
    \item The size distribution of our planetesimals formed close to the asteroid belt resembles the size distribution of primordial asteroids \citep{delbo2017}. Both can be described as a Gaussian distribution of combined planetesimal mass per planetesimal mass bin, which has its center at $125 km$ sized asteroids.
    \item All primordial asteroids (with the exception of Vesta and Flora) can be therefore be formed from the collapse of pebble clouds with an equivalent size of $152 km$.
    \item The "knee" at $100 km$ in the power law fits of initial planetesimal sizes corresponds to one standard deviation in mass lower than the most likely planetesimal mass of $125 km$. This means that transforming the Gaussian distribution centered at $125km$ of mass into cumulative number per size, a Knee at $100 km$ appears.
    }
\end{itemize}

Our numerical method facilitates studying the collisional evolution of collapsing pebble clouds and the formation of "solid" planetesimals in a consistent way and therefore makes two important claims: A: The mass distribution of asteroids and KBOs reflects the mass of the gravitationally unstable pebble cloud and can thus be explained by the underlying turbulence from for instance the streaming instability, and B: close binaries (for instance Patroclus and Menoetius which are going to be studied in the LUCY mission) and especially contact binaries like Arrokoth are direct outcomes of the multiple fragmentation of a pebble cloud into a planetesimal birth cluster. 

The outlook of this study has the following goals: extend the current simulations, test the stochasticity for all orbital radii, and increase the resolution. \responsetwo{The initial pebble cloud mass we use still has to be folded with an initial mass function of objects for particular clump masses and spins, which is beyond the scope of this paper.} Repeating each simulation with several different seeds, as done with the 1 AU case, will allow us to better constrain the initial mass and size distributions of planetesimals as well as the statistical spread of the binary and higher order system properties.

\responsetwo{We show that the mass function of the primordial asteroids can be interpreted as a normal distribution in mass, and our numerical simulations for the inner solar system give a very similar distribution. From that, we concluded that one can explain the observed mass distribution of asteroids as a result of stochastic fragmentation in the collapse process of a single mass pebble cloud. We could do this because in the inner solar system we could perform 12 simulations of the same cloud with randomized initial density perturbations. 

For our runs in the Kuiper Belt (beyond 30 AU) we were not able to do such an extensive study so far, because the simulation is significantly slower. We therefore only have a single run in the KB and refrain from over-interpreting this run as a general outcome for a size distribution. Once we have completed more runs for the KB, we can compare our synthetic mass function to observational data as well as try to fit it as a Gaussian distribution in mass, as we did for the asteroids, and likewise fit it to the exponentially tapered power laws in \citet{Kavelaars_2021} which find good agreement in the size distribution of cold classical KBOs. The size distribution in \citet{Kavelaars_2021} shows power-law behavior down to 20 km and hence appears in agreement with the IMF from streaming instability simulations \citep{Johansen2015_10.1126/sciadv.1500109,Simon_2016,Schafer2017A&A...597A..69S}.

How well a normal distribution in mass would fit the KBOs is still to be tested, yet the emphasis of normal distribution in mass vs. power law distribution in size and number differs. In the former, one derives a typical mass of planetesimals to use as an initial mass to study their further evolution, where the size slope towards smaller objects in the KBO distribution follows from collisional fragmentation (as in \citet{Marschall2022....164..167M}) rather than reflecting the initial size distribution. For the latter, the power law size distribution case, one directly tries to explain the KBO distribution as primordial, at least for the larger sizes. Therefore, both descriptions, Gaussian and power law, do not have to mutually exclude each other, but can be utilized for their purpose.
}

We are also interested in adding a finite elasticity to the forming planetesimals to better study their shape, which is currently not maintained in collisions. This could be done by using for instance a Tillotson EOS in combination with an elasticity model for icy and rocky material \citep{Schaefer2016}.

A long term N-body study of our planetesimal birth clusters sounds appealing, yet one should be aware that they cannot be considered to exist in isolation. At a local conversion rate of up to $10^{-4}$ $\mathrm{M_{\bigoplus}}$ per year and disk annulus \citep{Lenz2019,Lenz2020,Voelkel2020a} we will find many pebble cloud collapses to happen simultaneously. Plus, there is already the growing population of planetesimals and eventually embryos that would interact with the cluster.
Our simulation results can be fed into N-Body simulations as in \citet{MORBIDELLI2009558,Voelkel2020a,Voelkel2020b,Voelkel2022}. Thus, one can study the further evolution of planetesimals into planets and at the same time follow the fate of binary planetesimals through the history of our solar system.


\begin{acknowledgments}
This work is supported by the Deutsche Forschungsgemeinschaft (DFG, German Research Foundation) under Germany’s Excellence Strategy EXC 2181/1 - 390900948 (the Heidelberg STRUCTURES Excellence Cluster) and via priority program (DFG SPP) SPP 1833 "Building a Habitable Earth" under contracts: KL 1469/13-1 and KL 1469/13-2. Simulations were performed on the ISAAC cluster of the MPIA and the COBRA, HYDRA and DRACO clusters of the Max-Planck-Society, hosted at the Max-Planck Computing and Data Facility in Garching (Germany). B.P. is supported through a fellowship from the International Max Planck Research School for Astronomy and Cosmic Physics at the University of Heidelberg (IMPRS-HD).

Many thanks to Jesper Tjoa and Francesco Biscani for technical advice and discussion of our results. Alessandro Morbidelli and Marco Delbo were very helpful in comparing our numerical results to observational data. The authors are also very thankful to the referee for their insightful comments and questions, one of which led to the addition of an entirely new results section.

\end{acknowledgments}


\software{GIZMO \citep{10.1093/mnras/stv2180}, glnemo2 \citep{glnemo2}, scikit-learn \citep{scikit-learn}, yt \citep{2011ApJS..192....9T}, WebPlotDigitizer \citep{Rohatgi2020}}



\appendix
\section{Equation of State}
\label{section:EOS}
\begin{figure}
\centering
     \includegraphics[width=1.0\linewidth]{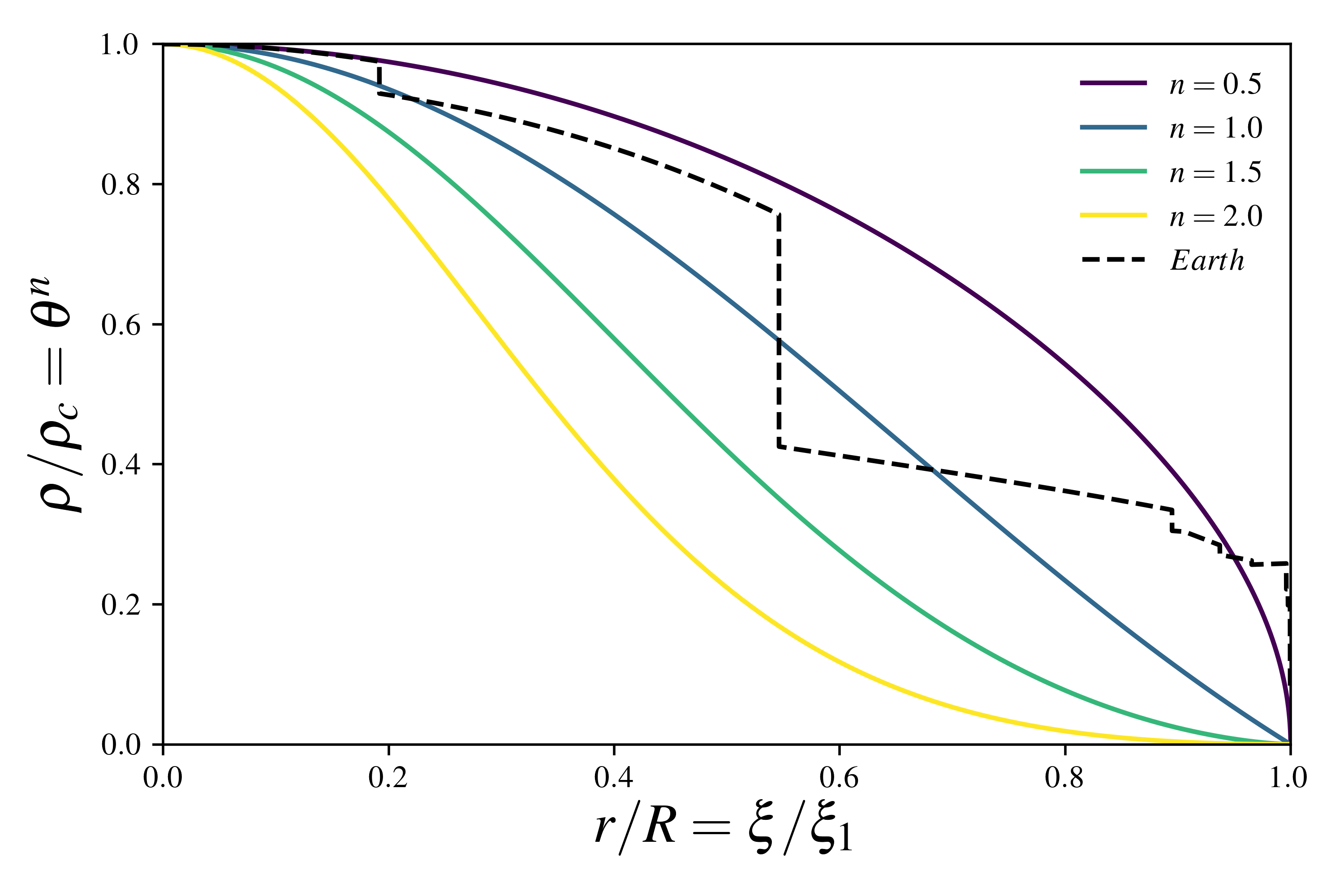}
    \caption{Dimensionless radial density profile of the Earth and various polytropes. Density and radius are normalized by central density and outer radius respectively.}
    \label{fig:polytrope}
\end{figure}

We use a polytropic EOS is of the form 
\begin{equation}
    P=K\rho^{1+\frac{1}{n}}
\end{equation}
where $n$ is the polytropic index, and $K$ is a constant related to entropy. The polytropic index determines the radial density profile of the material. We chose a polytropic index of $n=1$ because it best approximates the radial density profile of Earth, as shown in Figure \ref{fig:polytrope}. Moreover, we use a polytropic EOS because it allows us to choose the constant $K$ to set the hydrostatic equilibrium (HSE) radius of the collapsed cloud. We use this relation to set the HSE radius such that the mean final density of our planetesimals correspond to a physical value of $\rho\approx 1.0$ g/cc. 

The potential of a polytropic spherical cloud with self-gravity is described by the Lane-Emden equation. Solving the Lane-Emden equation provides a relationship between the pressure and density of our clouds. We use this solution to find a mathematical relationship between the polytropic constant and the HSE radius of our particle cloud. Below, we go through the solution to the Lane-Emden equation and derive the $K(\eta_c)$ relation.

We redefine density and pressure in terms of the central density $\rho_c$ and a dimensionless variable $\theta$ known as the polytropic temperature
\begin{equation}
    \rho=\rho_c \theta^n
\end{equation}
\begin{equation}
    P=K\rho_c^{1+\frac{1}{n}}\theta^{n+1}
\end{equation}
and redefine radius in terms of the length constant $\alpha$ and a dimensionless radius variable $\xi$
\begin{equation}
    r=\alpha \xi
\end{equation}
\begin{equation}
    \alpha^2 = \frac{(n+1)K\rho_c^{\frac{1}{n}-1}}{4\pi G}
\end{equation}

Solving for $K$, we get

\begin{equation}\label{eq:k_original}
    K=\frac{4\pi G}{(n+1)}\alpha^2\rho_c^{1-\frac{1}{n}}
\end{equation}

Our goal is to solve for the $K$ value that corresponds to the final radius of the formed planetesimals. To do this, we must find expressions for the equilibrium values of $\alpha$ and $\rho_c$. With these new variables, the Lane-Emden equation reads
\begin{equation}
    \frac{1}{\xi^2}\frac{d}{d\xi}\Big(\xi^2\frac{d\theta}{d\xi}\Big)=-\theta^2
\end{equation}

Let us define the ratio of the initial cloud radius to the collapsed equilibrium radius as the \textit{collapse ratio} $\eta_c$:
\begin{equation}\label{eq:eta}
    \eta_c \coloneqq \frac{R_f}{R_0}
\end{equation}

The radius of a polytrope is determined by solving the Lane-Emden equation for $\xi_1(n)$ which is the $\xi$ value where $\theta(n,\xi)$ first equals 0. Therefore, the equilibrium radius of our cloud is given by
 
\begin{equation}
    R_f=\alpha\xi_1(n)
\end{equation}
 
Using equation \ref{eq:eta}, we can write $\alpha$ in terms of the collapse ratio $\eta_c$ and the initial cloud radius $R_0$.
 
\begin{equation}\label{eq:alpha}
    \alpha=\frac{\eta_c R_0}{\xi_1(n)}
\end{equation}
 
Solutions to the Lane-Emden equation also provide the ratio of the central density to the average density $\frac{\rho_c}{\Bar{\rho}}(n)$. We can approximate the average density as $\Bar{\rho}=\frac{3M}{4\pi R_f^3}$, where $M$ is the total mass of our particle cloud. Since we initialize our particle cloud at a constant density of $\rho_{H^*}$, this gives a total mass of $M=\frac{4\pi}{3}R_0^3\rho_{H^*}$. The average density is then

\begin{equation}
    \Bar{\rho}=\Big(\frac{R_0}{R_f}\Big)^3\rho_{H^*}=\eta_c^{-3}\rho_{H^*}
\end{equation}

This gives the central density of our final clump in terms of the collapse ratio $\eta_c$ as

\begin{equation}\label{eq:rho_c}
    \rho_c=\Big(\frac{\rho_c}{\Bar{\rho}}(n)\Big)\rho_{H^*}\eta_c^{-3}
\end{equation}

Plugging equations \ref{eq:alpha} and \ref{eq:rho_c} into equation \ref{eq:k_original}, we get the following expression for $K$:

\begin{equation}\label{eq:K}
    K=\frac{4\pi G}{(n+1)}\Big(\frac{\eta_c R_0}{\xi_1(n)}\Big)^2\Bigg[\Big(\frac{\rho_c}{\Bar{\rho}}(n)\Big)\rho_{H^*}\eta_c^{-3}\Bigg]^{1-\frac{1}{n}}
\end{equation}

\begin{figure*}[htb]
\centering
    \includegraphics[width=1.0\textwidth]{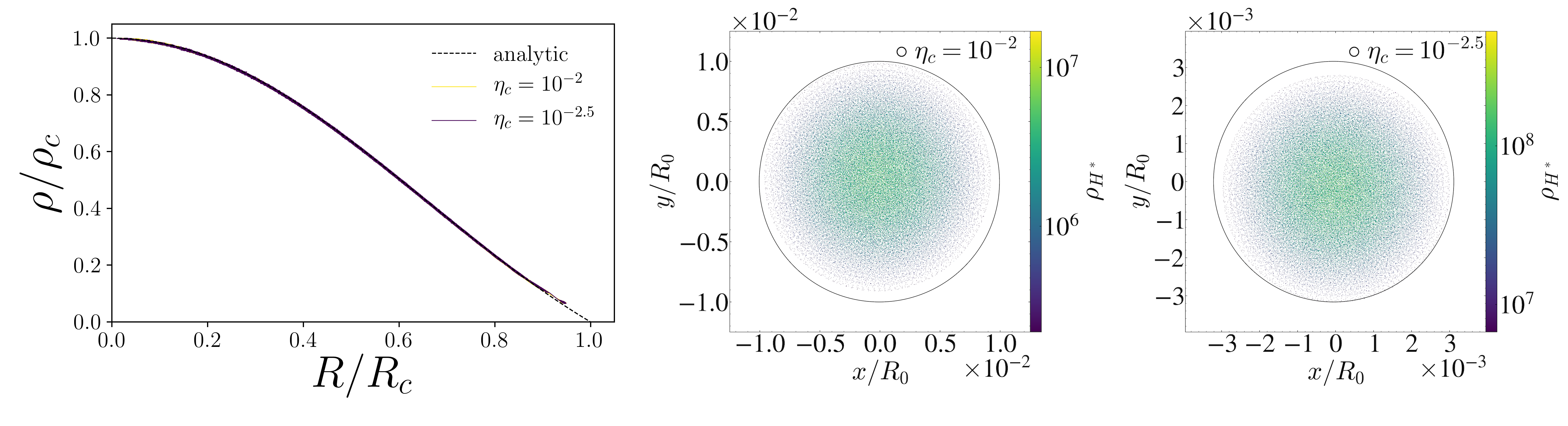}
    \caption{Tests validating our derived relation between the polytropic constant K and the collapse ratio $\eta_c$. The left plot shows the normalized radial density profile for collapsed clumps with different $\eta_c$ values compared to the analytic solution. The two plots on the left show both clumps compared to the analytically set final radius, indicated by the dotted circle.}
    \label{fig:collapse_test}
\end{figure*}

Figure \ref{fig:collapse_test} verifies the derivation of equation \ref{eq:k}.

\section{Clump Finding Routine}
\label{section:clumpfinder}

Here we give an overview of how the clustering algorithm DBSCAN works. DBSCAN takes two input parameters: the minimum neighbor radius for a point to be included in a cluster (\verb epsilon ) and the minimum number of points to create a cluster out of a group of points (\verb min_samples ). The steps of DBSCAN are:

\begin{enumerate}
    \item A random starting data point is picked, and the neighboring points within \verb epsilon  are found.
    \item If the number of neighbors within \verb epsilon  are at least \verb min_samples , then that data point becomes a core point. Otherwise, this point is labelled as noise.
    \item All neighboring points within \verb epsilon  of the core point are visited and marked as part of a cluster. Then neighbors of all the new cluster points within \verb epsilon  are also added to the cluster. This is repeated until all points in the \verb epsilon  neighborhood of the cluster points have been visited.
    \item Once the first cluster is complete, a new random starting point is selected and the process is repeated until all points are visited and labelled as noise or belonging to a cluster.
\end{enumerate}

First, we filter out noise by only passing points to DBSCAN that are above a minimum density $\rho_{min}=\%0.1\ \eta_c^{-3}\rho_{H^*}$. This means we define the edge of a clump as when the points reach 0.1\% of the central density. Considering our particles are n=1 polytropes, Figure \ref{fig:polytrope} indicates that this cutoff will include roughly 99.9\% of the final clump mass. To achieve generality across our simulations, we set the parameters \verb epsilon  according to the expected size the final planetesimals each simulation should form. Specifically, we set \verb epsilon  to be $\eta_c$ (predicted clump radius). We set \verb min_samples  to 100, meaning we only track clumps that are at least 0.1\% of the initial mass of the cloud. This gives the following 3 criteria for our clump finding routine:

\begin{enumerate}
    \item All points in a clump are $>=0.1\% \rho_c$
    \item Clump masses are $>=0.1\%\ M_{cloud}$
    \item All points in a clump are within a distance $\eta_c$ from at least one other clump point 
\end{enumerate}





\section{MFM vs. SPH}
\label{section:MFMvsSPH}

\begin{figure*}[htb]
\centering
    \includegraphics[width=\linewidth]{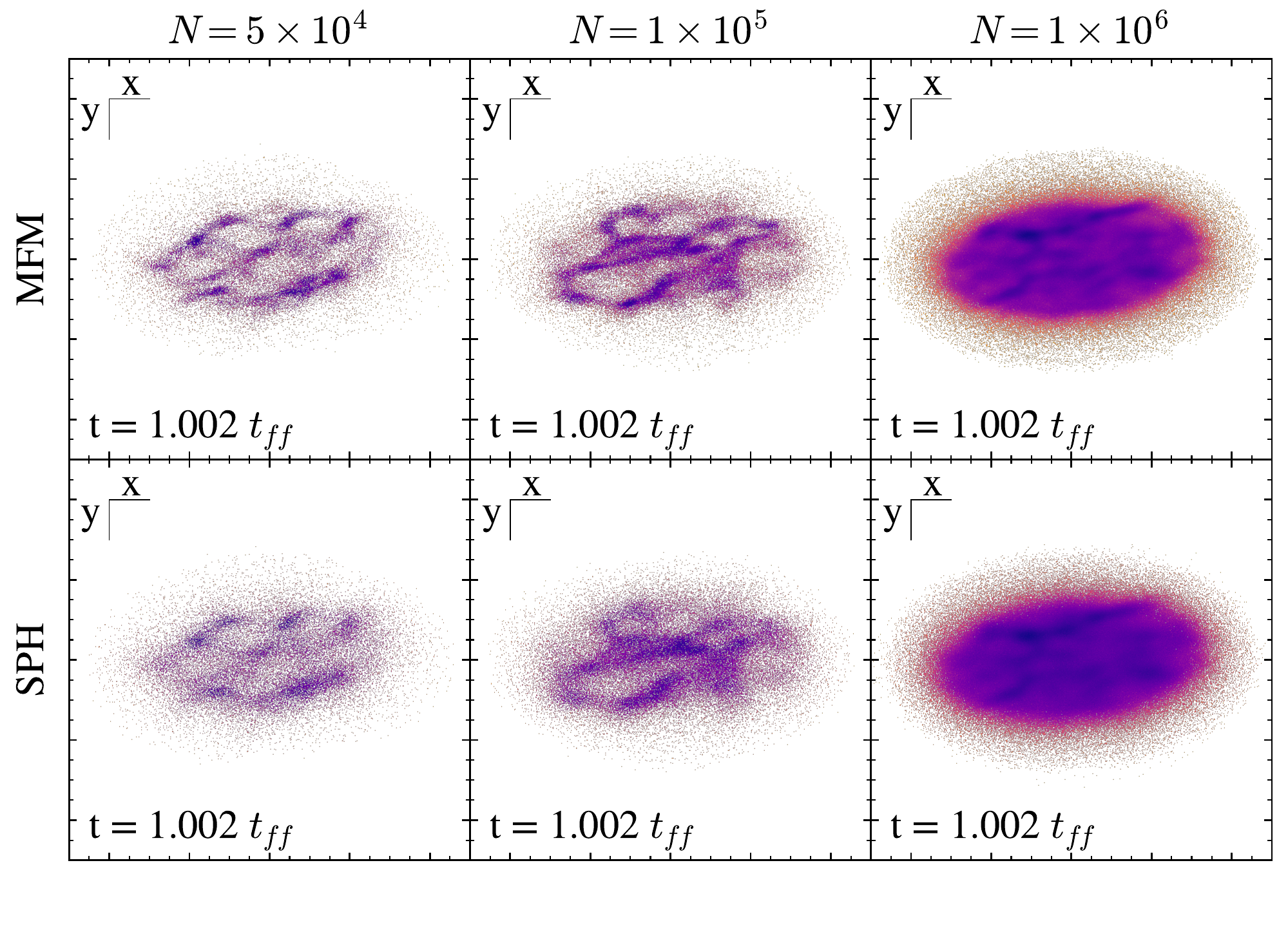}
    \caption{Comparison of the same collapse simulations done with the MFM and SPH hydro methods in GIZMO at various particle resolutions.}
    \label{fig:MFMvSPH}
\end{figure*}

The difference in shock preservation between SPH and MFM can be seen in practice for our setup in Figure \ref{fig:MFMvSPH}. The points of high density are visibly more defined in the MFM runs. 

\section{Resolution}
\label{section:resolution}

\begin{figure}[hbt]
\centering
    \includegraphics[width=\linewidth]{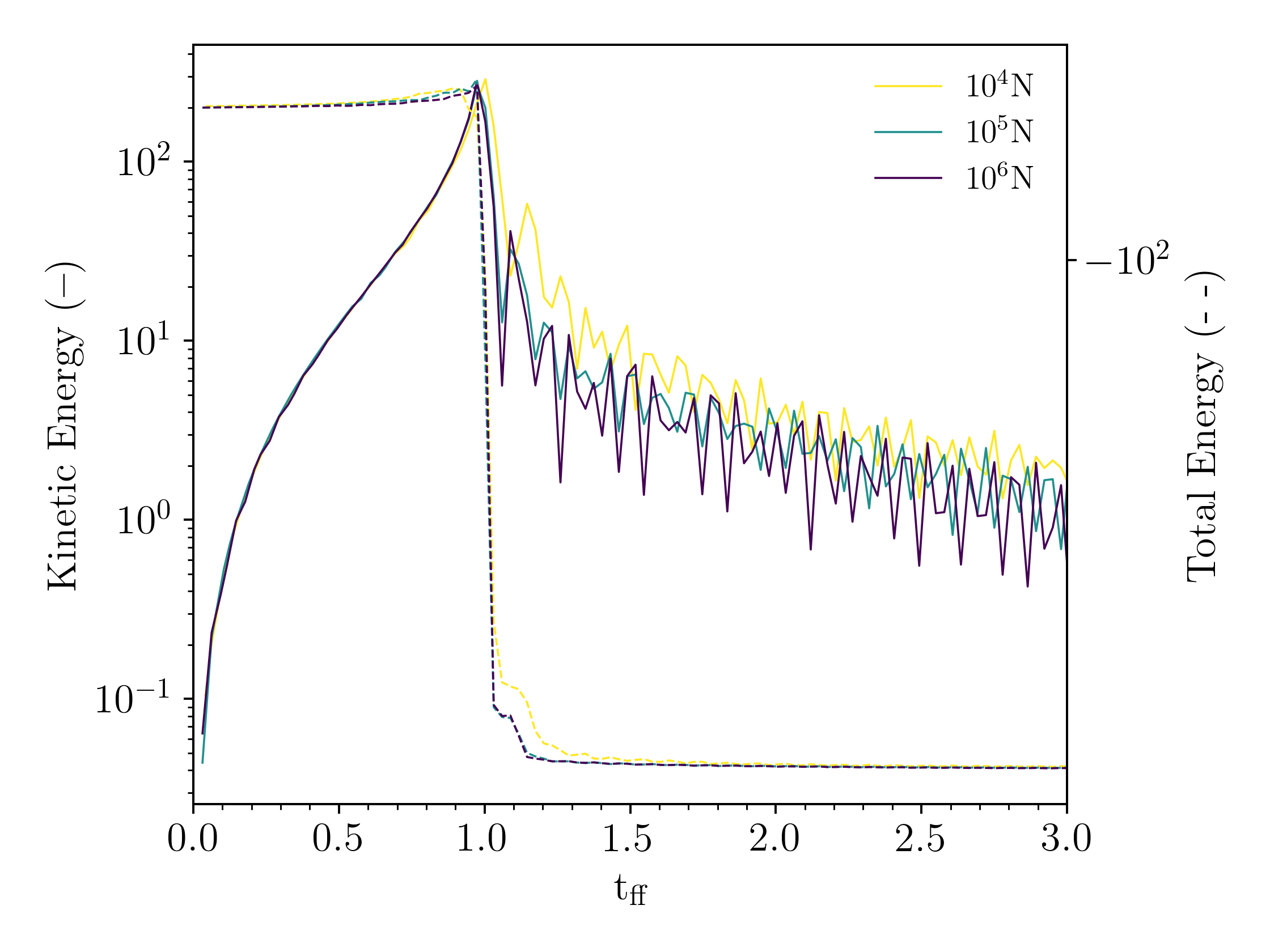}
    \caption{Kinetic (solid) and total (dashed) energy over time for a non-rotating collapse at various particle resolutions. Most of the system energy is dissipated during the shock at 1 free-fall time when the pressure starts to stop the gravitational collapse.}
    \label{fig:energyRes}
\end{figure}

\begin{figure}[hbt]
\centering
    \includegraphics[width=\linewidth]{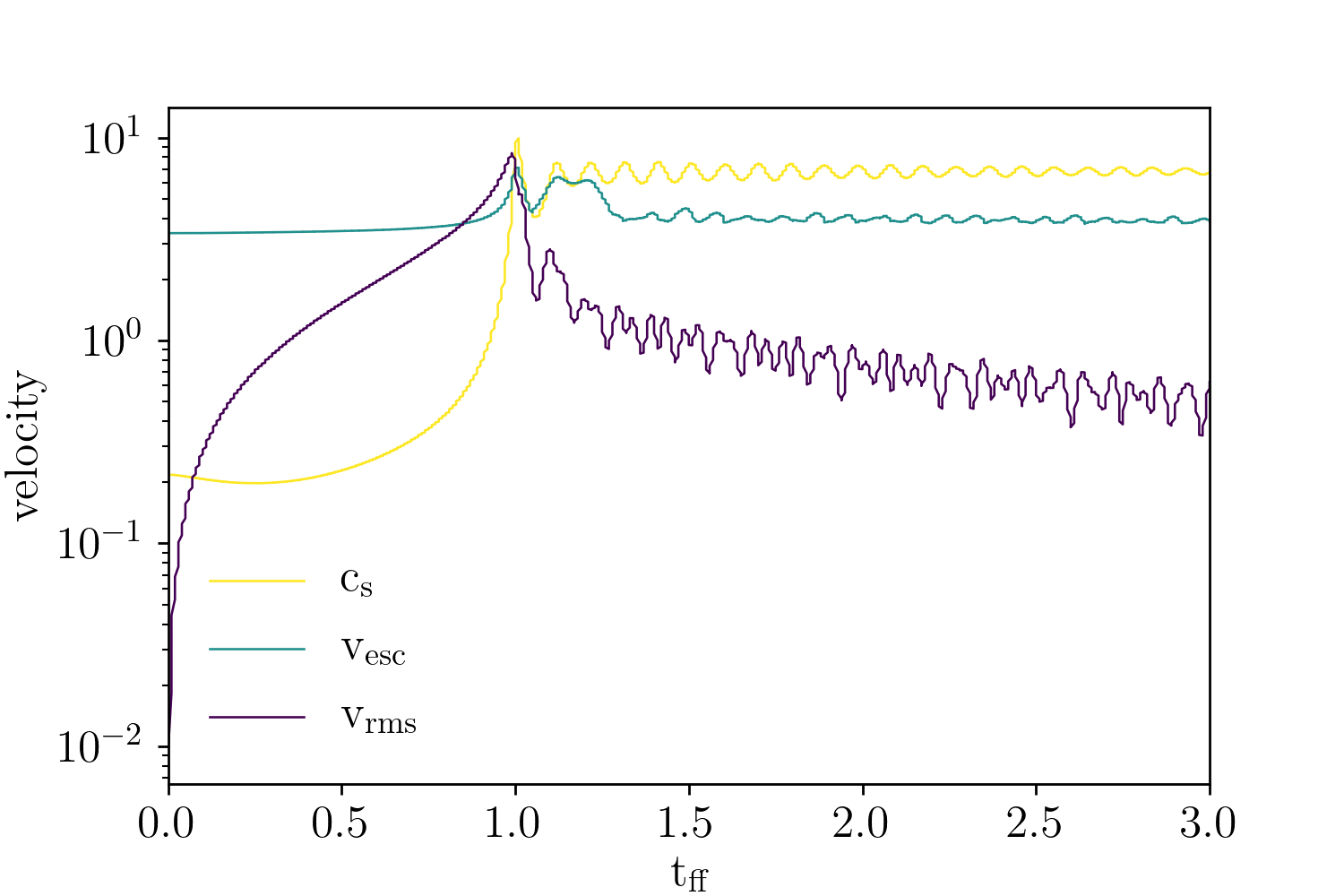}
    \caption{Time evolution of the sound speed, escape velocity, and RMS velocity of a non-rotating, spherically symmetric collapse simulation with time given in units of free-fall time. The energy dissipating shock is generated at 1 free-fall time when the rms velocities reach the sound speed of the material.}
    \label{fig:velTime}
\end{figure}

In a particle collision model with super-particles representing entire sub-clouds of particles, the resolution limit is set by whether the number of super-particles can reproduce the amount of kinetic energy dissipated through collisions in the physical system of the individual particles. Figure \ref{fig:energyRes} shows this dissipation in our model for increasing super-particle resolution. This test was done with a non-rotating collapsing uniform density sphere with 3 different numbers of particles. We used $10^5$ super-particles, as this is the maximum resolution computationally feasible for this study. This resolution is valid as we see a convergence in the kinetic energy dissipated for $10^5$ and $10^6$ super-particles. Figure \ref{fig:velTime} shows the sound speed of the non-rotating collapse at N=$10^5$ resolution over time, as well as the RMS and escape velocities for comparison. In the velocity plot, around 1 free-fall time is the point at which the collapse R.M.S. velocities go from sub-sonic to super-sonic, indicating that the particles have entered the pressure supported regime. 

\section{Centrifugal radius}
\label{app:RCent}
In contrast to \citet{nesvorny2019transneptunian}, we do not use pebble clouds extracted from streaming instability simulations, but we choose a simple initial condition of a spherical cloud to have a reproducible studies of the collapse process itself. 
The angular momentum content of these clouds corotating with the local Keplerian frequency $\Omega$ is 
\begin{equation}
    L = \frac{2}{5} m \Omega R_0^2.
    \label{Eq:L}
\end{equation}
This corresponds to two point masses rotating around each other at a distance of $R_1 = \sqrt{\frac{2}{5}} R_0$.
For centrifugal balance at $r_c$ they have to orbit with
\begin{equation}
    \omega^2 r_c = \frac{G m}{8 r_c^2}.
\end{equation}
Thus the corresponding angular momentum is
\begin{equation}
    L^2 = \frac{m^3 G}{8} r_c.
\end{equation}
Plugging in Eq.\ \ref{Eq:L} we get:
\begin{equation}
    r_c = \frac{4}{25 m} \Omega^2 R_0^4 \frac{8}{G} = R_0 \frac{32}{25} G \frac{M}{m} R_0^3 / R^3.
\end{equation}
With $R_0 = \frac{2}{3} a = \frac{2}{3} R \left( \frac{m}{3 M} \right)^{1/3}$ this leads to:
\begin{equation}
    r_c = R_0 \frac{32}{25}\frac{8}{27}\frac{1}{3} = 0.13 R_0.
\end{equation}
So any binary with a semi major axis less than $r_c$ needs some efficient re-distribution of angular momentum, leading to multiple fragmentation.

\bibliography{references}{}
\bibliographystyle{aasjournal}

\end{document}